\pdfoutput = 1 

\documentclass[pre,reprint,nofootinbib,twoside,unsortedaddress]{revtex4-1}	

\usepackage{bm}			

\usepackage{graphicx}		
\usepackage{amsmath, amssymb} 
\usepackage{datetime}		
\usepackage{wasysym}		

\usepackage{xspace}		

\newcommand{\ie}{\textit{i.e.,}\xspace}

\newcommand{\E}{\textrm{E}}
\newcommand{\BigOh}{O}
\newcommand{\Var}{\textrm{Var}}
\newcommand{\vect}[1]{\mathbf{#1}}

\newcommand{\Idc}{\ensuremath{I_\textrm{DC}}\xspace}

\newcommand{\MC}{\textrm{MC}\xspace}			
\newcommand{\FLs}{\textrm{IdS}\xspace}			
\newcommand{\FLc}{\textrm{channel}\xspace}			
\newcommand{\SJ}{\textrm{InS}\xspace}		%

\newcommand{\Na}{\textrm{Na}}
\newcommand{\K}{\textrm{K}}
\newcommand{\Nap}{$\textrm{Na}^+$}
\newcommand{\Kp}{$\textrm{K}^+$}

\newcommand{\chan}{\textrm{chan}}
\newcommand{\sub}{\textrm{sub}}

\begin{document}

\title{On stochastic differential equation models for ion channel noise in Hodgkin-Huxley neurons }
\date{\today}

\author{Joshua H. Goldwyn}
\affiliation{University of Washington, Department of Applied Mathematics}

\author{Nikita S. Imennov}
\affiliation{University of Washington, Department of Bioengineering}

\author{Michael Famulare}
\affiliation{University of Washington, Department of Physics}

\author{Eric Shea-Brown}
\affiliation{University of Washington, Department of Applied Mathematics}
\affiliation{University of Washington, Program in Neurobiology and Behavior}


\begin{abstract}

The random transitions of ion channels between conducting and non-conducting states generate a source of internal fluctuations in a neuron, known as \emph{channel noise}.  The standard method for modeling fluctuations in the states of ion channels uses continuous-time Markov chains nonlinearly coupled to a differential equation for voltage.  Beginning with the work of Fox and Lu \cite{Fox1994}, there have been attempts to generate simpler models that use stochastic differential equation (SDEs) to approximate the stochastic spiking activity produced by Markov chain models.  Recent numerical investigations, however, have raised doubts that SDE models can preserve the stochastic dynamics of Markov chain models \cite{Mino2002, Zeng2004,Bruce2007a,Bruce2009,Sengupta2010}.  

We analyze three SDE models that have been proposed as approximations to the Markov chain model: one that describes the states of the ion channels and two that describe the states of the ion channel subunits.  We show that the former \emph{channel-based} approach can capture the distribution of channel noise and its effect on spiking in a Hodgkin-Huxley neuron model to a degree not previously demonstrated, but the latter two \emph{subunit-based} approaches cannot.  Our analysis provides intuitive and mathematical explanations for why this is the case: the temporal correlation in the channel noise is determined by the combinatorics of bundling subunits into channels, and the subunit-based approaches do not correctly account for this structure.  Our study therefore confirms and elucidates the findings of previous numerical investigations of subunit-based SDE models.  Moreover, it presents the first evidence that Markov chain models of the nonlinear, stochastic dynamics of neural membranes can be accurately approximated by SDEs, even with as few as 60 \Nap and 18 \Kp channels.  This finding opens a door to future modeling work using SDE techniques to further illuminate the effects of ion channel fluctuations on electrically active cells.

\end{abstract}

\maketitle

\section{Introduction}    \label{sec:introduction}

Hodgkin and Huxley's mathematical model of action potentials dynamics \cite{Hodgkin1952} is a cornerstone of computational neuroscience.  This system of equations provides a \emph{conductance-based} framework for describing the dynamics of the membrane potential of a neuron.  The essential features of the Hodgkin-Huxley (HH) model are quantitative descriptions  of the permeability of a neuronal membrane to ion-specific currents (conductances) coupled to a current-balance equation that characterizes the voltage across a neural membrane. 
The physical basis for this empirical model is that conductances are determined by the proportion of ion channels in a \emph{conducting} (\ie an \emph{open} state), and that the states of the ion channels are determined by the configuration of components of the ion channels, referred to as \emph{subunits},  \emph{particles}, or \emph{gates} (see \cite[e.g.]{Dayan2001}).  The original HH equations can be interpreted as a model of the mean behavior of the ion channels and their subunits.  After the advent of techniques for recording the activity of individual ion channels \cite{Sakmann1995}, it was discovered that ion channels can transition between open and closed states in an apparently random manner and that this can generate a significant source of noise to the ionic conductances \cite{Hille2001}.  

This internal source of fluctuations is known as \emph{channel noise}, and is distinguished from sources of noise that are external to the neuron such as synaptic noise \cite[e.g.]{Destexhe2010} or noise that is present in an applied stimulus. 
Channel noise has important effects on neuronal dynamics and coding: it can alter the firing threshold \cite{Clay1983, White1998, Pecher1939}, spike timing \cite{Schneidman1998, Rossum2003}, interspike interval statistics \cite{Rowat2007}, the amount of stochastic resonance \cite{Schmid2001, Schmid2004}, and influence synaptic integration \cite{Cannon2010}. Channel noise can also contribute to the overall variability in the nervous system, which in turn may pose constraints on the fidelity of the motor and sensory systems of an animal \cite{White2000, Faisal2008, White1998, Balasubramanian2001, Polavieja2002} and limit neuron miniaturization \cite{Faisal2005}.

The classical HH formalism produces a deterministic description of neuronal dynamics, so alternative models have been proposed to account for channel noise.  These models assume that the activity of ion channels is governed by random transitions among a number of possible channel conformations, which leads to intrinsically \emph{stochastic} models of neuronal dynamics. Although a variety of models of this type have been proposed, including those that capture fractal properties of patch-clamp data \cite{Liebovitch2001} and history-dependence in the activity of ion channels \cite{Soudry2010}, the  most widely-used channel noise model is the Markov chain (\MC) model.  \MC models assume that the state of an ion channel is described by a discrete-state, continuous-time Markov chain, where each state in the chain represents a particular configuration of the ion channel.  The Markov property requires that a channel's transition from one state to the next depends on its current state alone, thus the transition rates are determined solely by the state of the channel and the voltage potential of the membrane.  As a consequence, all channels are coupled due to their common dependence on the membrane potential.  For a recent review of \MC models in computational neuroscience, see \cite{Groff2010}.

\MC models are an invaluable tool for investigating the effects of channel noise on neural dynamics and coding, but these models are computationally expensive to simulate and are difficult to analyze mathematically.  As a result, there has been widespread interest in formulating stochastic differential equation (SDE) models of channel noise.  This line of research was initiated by Fox and Lu \cite{Fox1994, Fox1997} and has been applied extensively in HH-type neuron models as well as models of calcium release from IP$_3$ receptors \cite{Shuai2002} (see \cite{Bruce2009} for a review of past applications of this approach).  The SDE model that is most commonly used extends the original HH equations by including noise terms in the differential equations that describe the gating variables. Computationally, this model can be orders of magnitude faster than \MC models \cite{Fox1997, Mino2002}, so it has often been used in place of, or as an approximation to, the \MC model.  Simulation studies have shown, however, that this SDE approach does not accurately replicate the stochastic response properties of the \MC models \cite{Mino2002,Zeng2004,Bruce2007a,Bruce2009,Sengupta2010} and it has been suggested that such models are inadequate for simulations of channel noise \cite{Faisal2010} or must be modified to correctly reflect the stochastic properties of the \MC models \cite{Bruce2009}. 

Despite recent indications that the commonly-used SDE model does not approximate the behavior of the \MC model, there has been no definitive study detailing the cause of discrepancies between the \MC and SDE approaches.  Moreover, other SDE models that have been proposed \cite{Fox1994, Shuai2002} have never been tested to gauge whether there may be alternative, and more accurate, reduced models of channel noise.  There are several possible reasons as to why an SDE model may not closely approximate a \MC model.  The system size expansion method for deriving an SDE model is an asymptotic result that is formally valid in the limit of a large number of channels; it is possible that there are too few ion channels in a realistic model neuron for these approximate methods to be accurate.  Another possible cause of the discrepancy between the two approaches is numerical error in the simulation algorithms \cite{Bruce2007a}.  Finally, it could be that the widely-used SDE models are formulated in a manner that neglects, or distorts, important dynamical and stochastic structure in the \MC model.  

In this paper, we will demonstrate that the formulation of the SDE is critical for preserving the stochastic character of \MC models.  In Section \ref{sec:HHintro}, we introduce three different SDE models that have been proposed in the literature.   Among these, we distinguish between \emph{channel-based} and \emph{subunit-based} SDE models and provide an intuitive explanation for why the channel-based approach is the more appropriate SDE framework.
We use a combination of mathematical analysis (Section \ref{sec:vclamp}) and simulation results (Section \ref{sec:Simulations}) to show
that the \MC model can, in principle, be well-approximated by a channel-based SDE model that was first introduced by Fox and Lu \cite{Fox1994}.  To our knowledge, ours is the first numerical implementation of the channel-based Fox and Lu model \cite{Fox:PersonalComm}.
Prior studies have provided numerical evidence that a widely-used subunit-based model does not accurately approximate the \MC model \cite{Mino2002,Zeng2004,Bruce2007a,Bruce2009,Sengupta2010},  
and our analysis confirms and elucidates these findings.
We conclude that properly defining the structure and dynamics of ion channels is critical to formulating SDE models in a way that is consistent with \MC models. 
We provide additional evidence for this conclusion by formulating reduced, quasistationary models based on our analytical results.  Simulations of these models show that temporal correlation in the noise, which is shaped by the structure of ion channels, is critical for accurately approximating responses of the \MC model.

\section{Conductance models based on ion channels and their subunits} \label{sec:HHintro}

We consider the HH model throughout this study, but our analysis is applicable to any conductance-based model with ion channels governed by linear, voltage-dependent kinetics. The membrane potential of an HH neuron is modeled as:
\begin{equation}
\label{eq_HH}
C \frac{dV}{dt} = -g_\Na(V-E_\Na) - g_\K(V-E_\K) -g_\textrm{L} (V-E_\textrm{L}) + I 
\end{equation}
where $C$ is the membrane capacitance, $E_\Na, E_\K$, and $E_\textrm{L}$ are reversal potentials for \Nap, \Kp, and leakage currents, respectively, and $I$ is the applied current.  Our central question is how to appropriately define the ion channel conductances ($g_\Na$ for sodium and $g_\K$ for potassium) when one wants to include channel noise. 
Generally, one defines the \Kp conductance as $g_\K = \bar g_\K f$, where $f$ is the fraction of \Kp channels that are open and $\bar g_\K$ is the maximal conductance per ion channel. Then the problem of appropriately reproducing \Kp channel behavior reduces to computing the evolution of $f$, the fraction of open channels.
In the following, we will describe a number of methods for computing $f$.  We outline the standard \MC model of ion channel kinetics \cite{Skaugen1979, Clay1983} and highlight how this approach relates to the classical (deterministic) HH model.  We will then consider three distinct approaches for defining SDE models: two that were first proposed by Fox and Lu \cite{Fox1994} and a variant suggested by Shuai and Jung \cite{Shuai2002}.

Capturing the kinetics of a single subunit is the starting point for all of the models considered here.  In the standard HH model, the \Kp channel has four independent identical subunits, traditionally given the symbol $n$, that must all be in an open state for the channel to be in the conducting state \cite{Fox1994, Dayan2001}.   
The kinetics of a individual subunits are described by a two-state process:
\begin{equation} \label{subunit_kinetics}
\mbox{Closed}  \overset{\alpha_n}{\underset{\beta_n}{\rightleftarrows}} \mbox{Open},
\end{equation}
where these subunits switch between open and closed states with the voltage-dependent transition rates \cite{Hodgkin1952}:
\begin{align}
\alpha_n(V)	&= \frac{0.01 (10-V)}{\exp((10-V)/10)-1}   \notag \\
\beta_n(V)	&=   0.125 \exp(-V/80).  \notag
\end{align}
To simplify the notation, we will often omit the explicit dependence on $V$ and write only $\alpha_n$ and $\beta_n$.

The \Nap channels are modeled using two different subunit types, traditionally labeled $m$ and $h$, each of which described by an open-closed kinetic scheme. 
The analysis of the two channel types is fundamentally the same, but entails significantly more notational complexity for the \Nap channel.  For conciseness, we will present detailed analysis of the \Kp channel, but report results for both channel types.

 The analysis of the \Nap channel can be carried out in an analogous fashion, but entails significantly more notational complexity, so we will present detailed analysis of the \Kp channel only.  We report analytical and simulation results for the both channels.
In the remainder of this section, we review how this building block of a two-state subunit has been used to construct models of the \Kp conductance.

\subsection{Markov Chain Ion Channel Model}
The kinetic scheme in Eqn.~\ref{subunit_kinetics} can be used to define a Markov chain that describes the behavior of a single subunit that randomly transitions between two states \cite[e.g.]{Dayan2001}.  If we let $p_\sub$ be the probability that the subunit is in the open state, then the evolution of this probability satisfies
\begin{equation} 
\label{eq_submaster}
\frac{dp_\sub}{dt} = \alpha_n (1-p_\sub) - \beta_n p_\sub.
\end{equation}
This equation follows from Eqn.~\ref{subunit_kinetics} and the fact that the probability that a subunit is closed is $1-p_\sub$.  Since the \Kp channel is assumed to consist of four statistically identical and independent subunits, its configuration can be modeled as a five-state Markov chain, where each state indicates the number of open subunits at a given instant in time:
\begin{equation}
\label{eq_KMarkovChain}
0  \overset{4\alpha_n}{\underset{\beta_n}{\rightleftarrows}} 1  \overset{3 \alpha_n}{\underset{2\beta_n}{\rightleftarrows}} 2  \overset{2\alpha_n}{\underset{3\beta_n}{\rightleftarrows}} 3 \overset{\alpha_n}{\underset{4\beta_n}{\rightleftarrows}} 4.
\end{equation}
The channel is said to be in the \emph{open} or \emph{conducting} state if all four subunits are open simultaneously.  
Let $\vect{p}$ be a column vector where the $i^{th}$ element represents the probability at time $t$ that a channel has $i$ open subunits, then this probability distribution evolves in time according to the master equation
\begin{equation} \label{eq_mcmaster}
\frac{d \vect{p}}{dt} = A \vect{p},
\end{equation}
where the matrix A is:
\begin{equation} \label{eq_Kdrift}
\left[ \begin{smallmatrix}
-4 \alpha_n & \beta_n & 0 &0 &0\\
4 \alpha_n & -(3\alpha_n + \beta_n)& 2\beta_n &0 &0\\
0 & 3 \alpha_n& -2(\alpha_n + \beta_n) & 3\beta_n &0\\
0 & 0 & 2\alpha_n &-(\alpha_n + 3\beta_n) & 4\beta_n \\
0 & 0 & 0& \alpha_n & -4\beta_n  \end{smallmatrix} \right] . \notag
\end{equation} 

The conductance for a population of \Kp channels is determined by the proportion of the channels in the open state, $g_\K = \bar g_\K f$, where $\bar g_\K$ is the conductance per \Kp channel and $f=N_o / N$ is the fraction of open \Kp channels, with $N_o$ is the number of open channels and $N$ is the total number of channels in the system.

\subsection{Deterministic Conductance Models} \label{subsec:ode}
If we consider an idealized neuron with an infinite number of statistically-identical and independent channels, we can obtain a deterministic description of the fraction of open channels $f$.  In this limit, the fraction of open channels is equivalent to the probability that any one channel will be open.   In other words, Eqn.~\ref{eq_mcmaster} also defines a deterministic model of conductance where $g_\K  = \bar g_\K p_4$ and $p_4$ is given by the solution of the system of ordinary differential equations in Eqn.~\ref{eq_mcmaster}.

At first glance, the deterministic definition of $g_\K$ appears to differ from that in the classical HH model, in which $g_\K=\bar g_\K n^4$.  As discussed in \cite{Dayan2001}, however, these two models are equivalent: first, note that $p_\sub$ in our notation can be identified with the gating variable $n$ in the HH model because both satisfy the differential equation \eqref{eq_submaster} and both represent the proportion of subunits that are open.  Next, observe that the entire system of differential equations in Eqn.~\ref{eq_mcmaster} can be derived from the single HH gating variable by making the following substitutions:
\begin{equation}
\label{eq_changevar}
p_i =  {4 \choose  i} (1-n)^{4-i} n^i, \qquad\text{where } i =  0, 1, 2, 3, \text{ or } 4.\notag
\end{equation}
For instance, setting $p_4 = n^4$, we find:
\begin{align}
\frac{d p_4}{dt}&=  \frac{d }{dt} [n^4] \notag \\
				&= 4 n^3 \frac{dn}{dt} \notag\\ 
				&= 4 n^3 \left[\alpha_n (1-n) - \beta_n n \right]  \notag\\
				&= \alpha_n p_3 - 4 \beta_n p_4 \notag
\end{align}
This equation is identical to the final row of Eqn.~\ref{eq_mcmaster} and the remaining equations in that system (as well as those for \Nap) can be derived in a similar manner.

\subsection{SDE Conductance Models}

\subsubsection{Channel SDE Model} \label{subsub:channel}
In the previous section we arrived at a deterministic model for \Kp conductance because we considered the case of infinitely many channels.  If we define the number of \Kp channels to be finite, however, we can derive stochastic models using a system-size expansion \cite{Gardiner2004}.  This method was first applied to the HH model by Fox and Lu \cite{Fox1994}.  Following their notation, we define $x_i$ to be the proportion of \Kp channels that have $i$ open subunits.  Since we are dealing with a finite population, the proportion of channels in a particular state, $x_i$, is no longer a measure of probability $p_i$.  Rather, the number of open subunits fluctuates from realization to realization, which inevitably leads to a stochastic description of the channel.  The system size expansion provides a formal method for deriving a SDE model based on the master equation \eqref{eq_mcmaster}.  Fox and Lu showed that the SDE for the \Kp channel is
\begin{equation} \label{eq_channelsde}
\frac{d \vect{x}}{dt} = 
 A \vect{x} + S \vect{\xi},
\end{equation}
where $\vect{x}$ is a vector of the $x_i$, $A$ is the matrix in Eqn.~\ref{eq_mcmaster}, $\vect{\xi}$ is a vector of five independent Gaussian white noise processes with zero mean and unit variance, and $S$ is the matrix square root of the diffusion matrix $D$,
\begin{widetext}
\begin{equation} \label{eq_diffusionmatrix}
D = \frac{1}{N}
\left[ \begin{smallmatrix}
4 \alpha_n x_0 + \beta_n x_1 & -(4 \alpha_n x_0 + \beta_n x_1)& 0 & 0 & 0 \\
-(4 \alpha_n x_0 + \beta_n x_1) & 4 \alpha_n x_0 + ( 3 \alpha_n + \beta_n ) x_1 + 2 \beta_n x_2& -(3 \alpha_n x_1 + 2 \beta_n x_2) & 0 & 0\\
0 & -(3 \alpha_n x_1 + 2 \beta_n x_2) & 3\alpha_n x_1 + 2 (\alpha_n + \beta_n) x_2 + 3 \beta_n x_3 &-(2\alpha_n x_2 + 3\beta_n x_3) & 0\\
0 & 0 & -(2\alpha_n x_2 + 3\beta_n x_3) &  2 \alpha_n x_2 + (\alpha_n + 3\beta_n)x_3 + 4\beta_n x_4& -(\alpha_n x_3 + 4\beta_n x_4)\\
0 & 0 & 0 & -(\alpha_n x_3 + 4\beta_n x_4) & \alpha_n x_3 + 4\beta_n x_4
\end{smallmatrix} \right],
\end{equation}
\end{widetext}
where $N$ is the number of channels. To our knowledge, neither Fox and Lu nor other researchers have implemented this channel-based SDE model \cite{Fox:PersonalComm}.  We will mathematically analyze it under voltage clamp conditions and perform numerical simulations to show that it accurately replicates the stochastic properties of the \MC model.  We refer to this model as the \emph{channel} SDE model because the variable $\vect{x}$ is defined based on the states of the ion channels.

\subsubsection{Subunit SDE Models} \label{subsub:subunit}
Unfortunately, the channel SDE model above does not preserve the dynamical structure of the classical HH equations:  closed states are distinguishable and must be modeled, expanding the dimensionality of the system significantly. In an attempt to avoid this increase in complexity, one can apply the system size expansion procedure to the subunits rather than to the states of the channels.  This leads to stochastic models that resemble the classical HH model, but include noise in the equations governing the subunit variables $m$, $n$, and $h$.   We refer to such approaches as \emph{subunit} SDE models.
The subunit approach leads to the following SDE for the proportion of open subunits $n$ \cite{Fox1994, Shuai2002}:
\begin{equation}
\label{eq_sdesubunit}
\frac{dn}{dt} = \alpha_n (1-n) - \beta_n n + \sigma_n(V) \xi(t), 
\end{equation}
where the stochastic term $\xi(t)$ is a Gaussian white noise process with zero mean and unit variance that is scaled in a voltage-dependent manner by $\sigma_n(V)$, where
\begin{equation}
\label{eq_sigman}
\sigma_n^2(V) = \frac{\alpha_n (1-n)+\beta_n n}{N}.
\end{equation}

The model for the channel population is then built from the subunit populations.
Since each \Kp channel is composed of four statistically identical and independent subunits, Shuai and Jung proposed a model in which the proportion of open \Kp channels is defined as the product of four independent realizations (denoted $n_i$) of solutions to the SDE in Eqn.~\ref{eq_sdesubunit} \cite{Shuai2002}. This defines the \Kp conductance to be $g_\K = \bar g_\K n_1 n_2 n_3 n_4$.  We refer to this as the \emph{independent subunit} model, \SJ. Shuai and Jung did not implement this method. Instead, they followed a method introduced by Fox and Lu \cite{Fox1994} in which only one realization of a solution to Eqn.~\ref{eq_sdesubunit} is computed (denoted $n$), and this realization is raised to the fourth power. This defines the \Kp conductance to be  $g_\K = \bar g_\K n^4$, built out of four identical subunit populations.  We refer to this as the \emph{identical subunit} model, \FLs.  In the limit of an infinite number of \Kp channels, both of the subunit models converge to the deterministic HH model.

\subsection{The Distinction Between Subunit and Channel Models}
\label{subsec:intuitive}
The fundamental difference between the channel SDE model in Section \ref{subsub:channel} and the subunit SDE models in Section \ref{subsub:subunit} is that in the former, one first groups subunits together to construct a channel and then defines the dynamics of the proportion of channels in each state.  In the latter, one defines the dynamics of subunits first, before grouping the subunits together to compute the conductance of the channel.  We note that the deterministic model in Section \ref{subsec:ode} derived from the master equation is a channel-based approach while the classical HH model is a subunit-based approach. Nonetheless, as discussed above, the two models are equivalent.  It is tempting, therefore, to conclude that both the channel and subunit SDEs will also produce identical stochastic models.  As we will show in the remainder of this study, these two approaches generate distinct stochastic processes: channel-centric SDE models can approximate the channel noise and spiking statistics of the \MC model, but the subunit-based SDEs cannot.

\begin{figure}[!tbp]
\includegraphics[width=8.6cm]{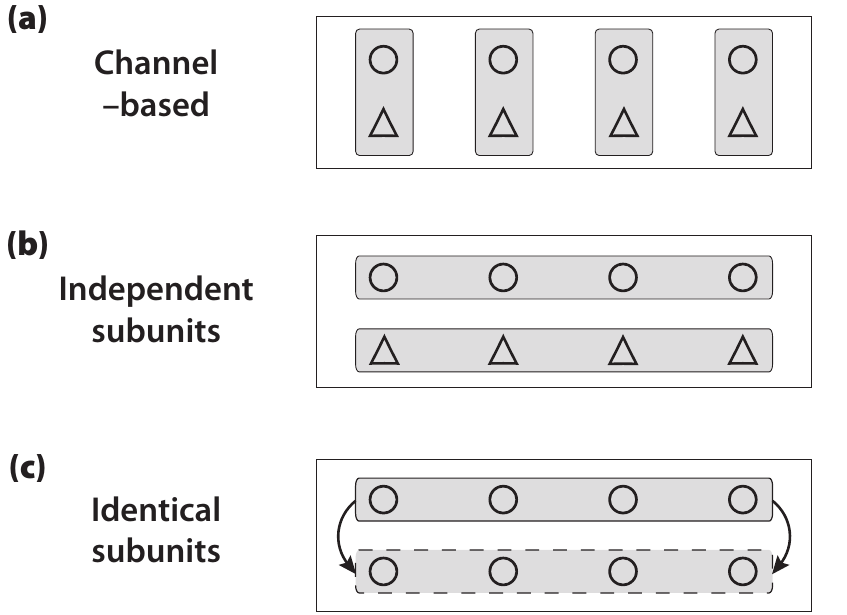}
\caption{Illustration of conceptual differences between channel-based and subunit-based models.  In this example, each channel consists of two subunits: ``$\bigtriangleup$'' and ``\Circle''.  \textbf{(a)} In the channel-based model, subunits are first grouped together to form channels (vertical rectangles) and the ionic conductance is determined by the fraction of channels in the conducting state.  \textbf{(b)} In the independent subunits (\FLs) approach, the subunits are divided into two classes (horizontal rectangles) and the fraction of open subunits is computed by averaging over all subunits in each class.  The proportions of open subunits in each class are then used to approximate the fraction of  channels in the conducting state.  \textbf{(c)}  Identical subunit (\SJ) models assume both subunits classes are identical.}
\label{fig:diagram}
\end{figure}

To gain some intuition for how the subunit and channel SDE approaches differ, consider the following example of a neuron with $N$ channels, where each channel consists of two statistically identical and independent subunits.  This configuration is illustrated in panel (a) of Fig. \ref{fig:diagram}. The analysis can be extended to the four subunit \Kp channel, but for illustrative purposes we consider the simpler case of two subunits.  At a given instant in time, define the state of the $i^{th}$ subunit in each class by the binary random variables $z_{i_1}$ and $z_{i_2}$.  These variables take the value of one with probability $p_\sub$ and are zero otherwise.  The probability that the $i^{th}$ channel is open is determined by the probability that both subunits are open, $p_{sub}^2$.  A channel-based approach defines the conductance from the proportion of open channels, so we average over all channels to obtain the proportion of open channels:
\begin{equation}
\label{eq_fchannel}
f_\chan= \frac{1}{N} \sum_{i=1}^N z_{i,1} z_{i,2}
\end{equation}
The $z_{i,1}z_{i,2}$ are themselves identical and independent binary random variables that take the value one with probability $p_\sub^2$, thus the mean and variance of $f_\chan$ are:
\begin{align}
 E[f_\chan] &= p_\sub^2 \notag \\
 \Var[f_\chan] &= \frac{1}{N} p_\sub^2 (1-p_\sub^2) \notag
\end{align}

%
%
%

The subunit approaches do not begin by grouping subunits into predefined channels.  Instead, one first computes the fraction of open subunits by averaging over each class of subunits, as shown in panel (b) of Fig. \ref{fig:diagram}.  The proportions of open subunits in the two subunit classes are:
\begin{equation}
\label{eq_fsub}
f_{\sub,j}= \frac{1}{N} \sum_{i=1}^N z_{i,j}  \mbox{ where } j = 1 \mbox{ or } 2. \notag
\end{equation}
If we assume that each subunit class is independent and define the proportion of open channels $\tilde f_\chan$ to be the product of the $f_{\sub,1}$ and $f_{\sub,2}$, then we can write the proportion of open channels as:
\begin{equation}
\tilde f_\chan = \frac{1}{N^2} \sum_{i,j=1}^N z_{i,1}z_{j,2}. \notag
\end{equation}
The proportion of open channels under the subunit approach is therefore an average of $N^2$ binary random variables rather than $N$ random variables, as in Eqn.~\ref{eq_fchannel} for the channel approach.
The probability that the product $z_{i,1}z_{j,2}$ is equal to one is $p_\sub^2$, so the expected value of $\E [\tilde f_\chan] = p_\sub^2$, identical to  $\E [f_\chan]$.  The variance in the two models, however is different.  To see this, write the variance of $\tilde f_\chan$ as the sum of covariances:
\begin{equation}
\label{eq_sumvar}
\Var [\tilde f_\chan] = \frac{1}{N^4} \sum_{i,j,k,l=1}^N \mbox{Cov}(z_{i,1}z_{j,2}, z_{k,1}z_{l,2})  \notag
\end{equation}
This sum is over all of the $N^4$ possible pairings of $z_{i,1}, z_{j,2}, z_{k,1}, \mbox{ and } z_{l,2}$.   To leading order in $N$, the dominant contribution to this sum is among pairings that have one index in common. There are $N^2$ possible pairs of $z_{i,1}$ and $z_{j,2}$, and for any given pair there are $2(N-1)$ ways to choose the indices of $z_{k,1}$ and $z_{l,2}$ such that $i=k$ or $j=l$.  To leading order in $N$, therefore, there are $2N^3$ of these terms, and the variance of $\tilde f_\chan$ can thus be written as:
\begin{equation}
\label{eq_vartildefchan_sj}
\Var[\tilde f_\chan ]= 2 \rho \Var [f_\chan]\quad,
\end{equation}
where $\rho$ is the correlation coefficient between random variables of the form $z_{i,1}z_{j,2}$ and $z_{j,1}z_{l,2}$ where either $i=k$ or $j=l$.
A straightforward calculation shows that $\rho = \frac{p_\sub}{1+p_\sub}$.  Since $p_\sub$ takes values in $[0,1]$, $\rho$ is bounded between zero and one half.  Equation \ref{eq_vartildefchan_sj} implies, therefore, that the variance of the fraction of open channels using the subunit-based averaging is always smaller than in the channel-based approach.  Moreover, the variance decreases with $p_\sub$.  This implies that the subunit method underestimates the variance when $p_\sub$ is small.

An assumption in the above analysis is that there are two independent classes of subunits.  This can be thought of as the analogue of the \SJ SDE model discussed above.  Another variation of the subunit approach, used in the \FLs SDE, is to assume that the two subunit classes are identical.  This approach is illustrated in panel c of Fig. \ref{fig:diagram}.  Since both subunit classes are identical, we drop the second subscript and define a random variable $z_i$ that represents the state of the $i^{th}$ subunit in both classes.  The proportion of open channels can then be rewritten as:
\begin{equation}
\tilde f_\chan = \frac{1}{N} \sum_{i=1}^N z_{i}^2. \notag
\end{equation}
The expected value of $\tilde f_\chan$, which is given by the probability that $z_i^2=1$, is identical to $\E [f_\chan]$.  The variance for this approach is:
\begin{equation}
\label{eq_sumvar_fl}
\Var [ \tilde f_\chan] = \frac{1}{N^4} \sum_{i,j,k,l=1}^N \mbox{Cov}(z_{i}z_{j}, z_{k}z_{l}).  \notag
\end{equation}
As above, to leading order in $N$ it suffices to consider the covariance of pairings that share one index in common.  There is no distinction between the two subunit classes, so $\mbox{Cov}(z_{i}z_{j}, z_{i}z_{k}) = \mbox{Cov}(z_{i}z_{j}, z_{k}z_{i})$.  There are therefore twice as many such terms as in the previous approach, and we find:
\begin{equation}
\label{eq_vartildefchan_fl}
\Var[\tilde f_\chan ]= 4 \rho \Var [f_\chan]
\end{equation}
with the same correlation coefficient  $\rho = \frac{p_\sub}{1+p_\sub}$.

This analysis illustrates how averaging across channels and averaging across subunits leads to fundamentally different probabilistic descriptions of the proportion of open channels.  In particular, since $\frac{p_\sub}{1+p_\sub} \le \frac{1}{2}$,  Eqn.~\ref{eq_vartildefchan_sj} guarantees that the variance of the proportion of open channels given by the subunit model with two independent classes of subunits will never exceed the variance given by the \MC model.  Equation \ref{eq_vartildefchan_fl} shows that the variance in the subunit model with identical subunit classes will always be twice as large as the variance in the \FLs model.  Depending on whether $\frac{p_\sub}{1+p_\sub}$ is smaller or larger than one fourth, therefore, the variance in the subunit model with identical subunit classes can be either smaller or larger than the variance of the \MC model.
These differences are a direct consequence of how each approach aggregates the channels' subunits.  Importantly, we observe that the differences between these approaches will persist for any finite number of channels.  In the limit of infinitely many channels, the variance goes to zero, so all of the modeling approaches discussed here become equivalent.
It is a straightforward exercise to extend this analysis to the case of four subunits (\ie the \Kp channel), and a similar discrepancy between the channel and subunit approaches holds in that case.  

The three methods for grouping subunits that we have considered represent the three different approaches to performing a system-size expansion that have been proposed by Fox and Lu \cite{Fox1994, Fox1997} and Shuai and Jung \cite{Shuai2002}.  These combinatorial arguments provide an intuitive understanding for why the three SDE approaches that we are studying will lead to channel noise models with different statistical properties.  We now confirm this by directly analyzing the SDE and \MC models.

\section{Voltage Clamp Analysis of Stochastic Models} \label{sec:vclamp}

\subsection{Stationary Distribution}

We now analyze the stochastic character of each of these SDE models and compare them to the \MC model. For the purposes of modeling voltage potential across the membrane, we are interested in the conductance of the ion channels so we seek to characterize the probability distribution of the fraction of open channels, which we denote $f$.  To simplify the analysis, we will mimic the experimental technique of voltage clamp and perform our analysis while holding the membrane potential constant.  

\subsubsection{Markov Chain Model}
\begin{figure}[!tbp]
\includegraphics[width=8.6cm]{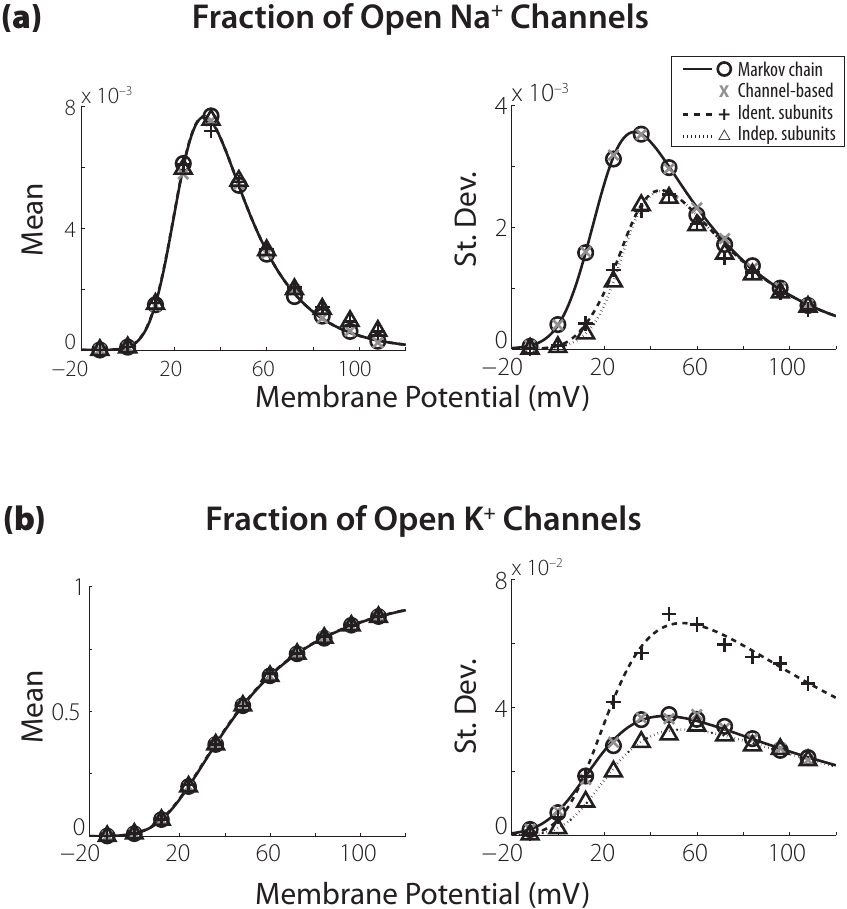}
\caption{Analytical (lines) and numerical (symbols) results for means and standard deviations of the fractions of open channels in voltage clamp.  The membrane area is 10 $\mu m^2$ (600 \Nap channels and 180 \Kp channels).  The abscissa gives the voltage clamp value.  \textbf{(a)} Results for \Nap channels.  \textbf{(b)} Results for \Kp channels.  Analytical results for the channel SDE model are not shown because they are identical to those of the \MC model.}
\label{fig:vclamp_mean_std}
\end{figure}

In the \MC model, each channel consists of four subunits that transition between open and closed states.  In voltage clamp this process is homogeneous in time so the stationary distribution for the number of open channels can be completely determined \cite{Gardiner2004}.  We are primarily interested in the stationary probability that all four subunits are open because this is equivalent to the probability that the channel itself is open.  From Eqn.~\ref{eq_submaster}, the equilibrium value of $p_\sub$ is $\frac{\alpha_n}{\alpha_n + \beta_n}$.  The probability that the \Kp channel is open is therefore:
\begin{equation}
p_\chan  = \frac{\alpha_n^4}{(\alpha_n+\beta_n)^4}. \notag
\end{equation}
All \Kp channels are assumed to be statistically identical and independent, so the distribution of the total number of open \Kp channels at a given time is a binomial distribution with population parameter $N$ (the total number of \Kp channels) and bias parameter $p_\chan$.  
To define the distribution of the fraction of open channels $f$, we rescale the binomial distribution by $1/N$.  
The mean and the variance of this stationary distribution are:
\begin{align}
\E_{\MC}[f]		&= p_\chan &\triangleq \mu_\chan	\label{EMC} \\
\Var_{\MC}[f]   &= \frac{p_\chan (1-p_\chan)}{N} &\triangleq \sigma_\chan^2	 \label{VMC},
\end{align}
which we define as $\mu_\chan$ and $\sigma^2_\chan$, respectively, as a shorthand.  Note that these quantities are functions of $V$, even though we do not explicitly include this dependence in our notation.  Equation \ref{EMC} shows that the mean does not depend on the number of channels and Eqn.~\ref{VMC} shows that the variance scales with $1/N$.  The mean and standard deviation of the fraction of open channels are shown by the solid black lines in Fig. \ref{fig:vclamp_mean_std}.  The mean and variance for the \Nap channel can be computed in a similar fashion and Fig. \ref{fig:vclamp_mean_std} includes those results.  The results shown are for a membrane area of 10~$\mu m^2$, which corresponds to 180 \Kp and 600 \Nap channels.

%
%
%

\subsubsection{Channel SDE Model}
To analyze the channel SDE model (Eqn.~\ref{eq_channelsde}), we apply the simplification suggested by Fox and Lu: we set values of the state variables in the diffusion matrix (Eqn.~\ref{eq_diffusionmatrix}) to their mean equilibrium values.  We refer to this approximation as the \emph{equilibrium noise approximation} and show in Appendix \ref{appendix:noise} that, for the voltage clamp case, it holds to $\BigOh(1/N^2)$.  In principle, the values of each $x_i$ should be confined to $[0,1]$ since they represent proportions of open channels, but to simplify the mathematical analysis we do not impose this condition. Under these simplifications, the SDE model is a multivariate Ornstein-Uhlenbeck (OU) process that by definition has a Gaussian-distributed stationary distribution.  The mean and variance of the stationary process can be calculated directly using standard methods \cite{Gardiner2004}. We find that the stationary distribution of the fraction of open \Kp channels is given by:
\begin{equation}
\label{ExactSS}
p(f) = \frac{1}{\sqrt{2 \pi \sigma_\chan^2}}e^{-\frac{(f-\mu_\chan)^2}{2 \sigma_\chan^2 }}. 
\end{equation}
This is the expected result for a system size expansion since, in the limit of a large number of channels, the binomial distribution for the \MC model can be approximated by a Gaussian distribution with mean $\mu_\chan$ and variance $\sigma_\chan^2$.  Any description of channel noise that aims to reproduce the behavior of the Markov state model should have this as its limiting distribution, so this result is our first confirmation that the channel SDE model provides an accurate approximation to the \MC model.

\subsubsection{Subunit SDE Models}
To analyze the subunit-based SDE models, we apply the same approximations: we replace $n$ in the equation for $\sigma_n^2$ (Eqn.~\ref{eq_sigman}) with its mean value at  equilibrium.  Thus, we replace $\sigma_n^2$ with $\overline{\sigma}_n^2 = \frac{2 \alpha_n \beta_n}{N(\alpha_n+\beta_n)}$ and do not restrict the values of $n$ to the interval $[0,1]$.  As was the case for the channel SDE, these approximations allow us to rewrite the SDE for $n$ (Eqn.~\ref{eq_sdesubunit}) as an OU process:
\begin{equation}
\label{eq_nou}
\frac{dn}{dt} = \frac{1}{\tau_n} (\mu_\sub - n) + \overline{\sigma}_n \xi_t,
\end{equation}
where  $\tau_n = \frac{1}{\alpha_n+\beta_n}$ and $\mu_\sub = \frac{\alpha_n}{\alpha_n+\beta_n}$.  The stationary distribution of $n$ (the fraction of open subunits) is therefore Gaussian-distributed with mean $\mu_\sub$ and variance $\sigma_\sub^2 = \frac{\alpha_n \beta_n}{N(\alpha_n+\beta_n)^2}$.  Note that $\sigma_\sub^2$ is scaled by $1/N$; for simplicity we report  analytical results to $\BigOh(1/N)$.

In Section \ref{subsub:subunit} we discussed two methods for defining the proportion of open \Kp channels based on the stochastic dynamics of the subunits.  If we follow the approach of Shuai and Jung and combine four statistically identical and independent solutions to Eqn.~\ref{eq_nou}, the stationary distribution for the proportion of open channels is defined by the product of four independent and identically distributed Gaussian random variables:
\begin{equation} \label{PSJ}
p_{\SJ}(f)\!=\!\displaystyle \int\!dn_1\ldots dn_4 \left[\prod_{j=1}^4 \frac{e^{-\frac{(n_j-\mu_\sub)^2}{2\sigma_\sub^2}}}{\sqrt{2\pi\sigma_\sub^2}}\right]\!\delta\!\left(\!f-\prod_{j=1}^4n_j\!\right)\!. \notag
\end{equation}
Unlike the channel SDE distribution in Eqn.~\ref{ExactSS}, in the limit of a large number of channels, this distribution does not approach a Gaussian.  This model, therefore, is fundamentally incompatible with the \MC model.  Furthermore, it is straightforward to compute the mean and variance of this distribution directly from the first two moments of the subunit distribution.  We find:
\begin{align}
\E_{\SJ}[f]		&= \mu_\sub^4 \label{ESJ}  \\
				&= \mu_\chan \notag\\
\Var_{\SJ}[f]	&=4\mu_\sub^6\sigma_\sub^2 + \BigOh(N^{-2}) \label{VSJ}  \\
				&\neq	\sigma_\chan^2 . \notag
\end{align}
We note that this leading order result for the moments can also be obtained following the combinatorial approach outlined in Section \ref{subsec:intuitive}.

If we go further and assume that all four populations of subunits are identical and perfectly correlated (\ie following the approximation proposed by Fox and Lu), then the stationary distribution for the fraction of open \Kp channels is given by the Gaussian distribution for a single subunit raised to the fourth power.  The distribution in this case has the closed form:
\begin{equation}
p_{\FLs}(f)=\frac{1}{4\sqrt{2\pi\sigma_\sub^2}}f^{-3/4}e^{-\frac{(\sqrt[4]{f}-\mu_\sub)^2}{2\sigma_\sub^2}}. \notag
\end{equation}
This distribution also does not limit to a Gaussian and is fundamentally incompatible with the \MC model.
As above, we can compute the mean and variance of $n^4$:
\begin{align}			
\E_{\FLs}[f] 	&=\mu_\sub^4+6 \mu_\sub^2\sigma_\sub^2 + \BigOh(N^{-2}) \label{EFL}\\
				&=\mu_\chan +\BigOh(N^{-1}),  \notag \\
\Var_{\FLs}[f] 	&=16\mu_\sub^6\sigma_\sub^2 + \BigOh(N^{-2})  \label{VFL}\\
				&\neq  \sigma_\chan^2 . \notag
\end{align}

Equations~\ref{ESJ} and \ref{EFL}, show that the mean fraction of open channels computed with the subunit approaches agrees with the \MC model in the limit of a large number of channels.
  The variance, however, is poorly described.  
For instance, the ratio of the open channel variance of the \MC model (Eqn.~\ref{VMC}) to that of the \FLs model (Eqn.~\ref{VFL}) is:
\begin{equation}
\frac{\sigma_\chan^2}{\Var_{\FLs}(f)}=\frac{1+\mu_\sub+\mu_\sub^2+\mu_\sub^3}{16\mu_\sub^3}. \notag
\end{equation}
For subthreshold values of $V$,  $\mu_\sub$ is small and therefore the \FLs model drastically underestimates the magnitude of the channel noise.  

These analytical results for the two subunit SDE models are plotted in Fig. \ref{fig:vclamp_mean_std} with dotted (independent subunit populations; \SJ) and dashed (identical subunit populations; \FLs) lines.  
Note that the channel noise in the \Nap channel is also underestimated by both subunit models for $V$ near the resting potential of zero millivolt.

\subsection{Autocorrelation in Voltage Clamp} \label{autocorrelation}

We now analyze temporal correlations in the proportion of open channels for a given voltage clamp level.  As in prior sections, equations presented here depend on voltage potential $V$, but to simplify notation we do not explicitly indicate this.
If we denote the time series of the proportion of open channels as $f(t)$, then the autocorrelation function for $f(t)$ is:
\begin{equation}
\label{eq_autocorr}
R(t)  = \frac{\mbox{E}\left[ f(t)f(0)\right]-\E[f(0)]^2}{\Var[f(0)]}. \notag
\end{equation}
We assume $R(t)$ does not depend on the initial time since our analysis is restricted to the stationary distribution of open channels in voltage clamp. 

\subsubsection{Markov Chain Model}
\begin{figure}[!tbp]
\includegraphics[width=8.6cm]{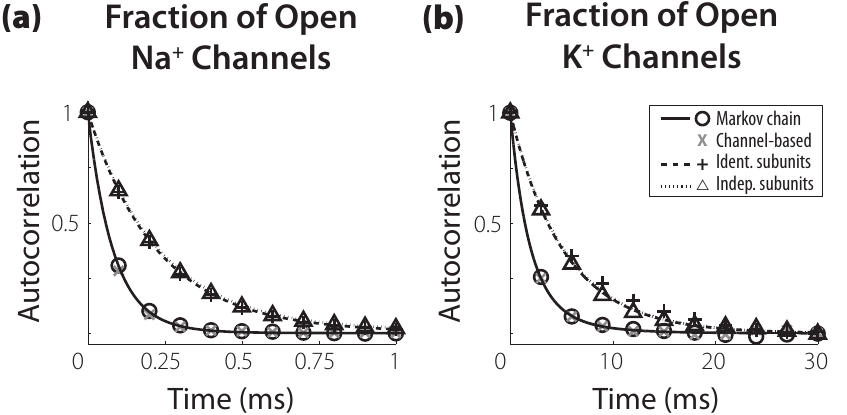}
\caption{Analytical (lines) and numerical (symbols) results for autocorrelations $R(t)$ of the fractions of open channels in voltage clamp.  The voltage clamp is set to 0~mV and the membrane area is 10~$\mu m^2$ (600 \Nap channels and 180 \Kp channels).  \textbf{(a)} Results for \Nap channels.  \textbf{(b)} Results for \Kp channels.  Analytical results for the channel SDE model are not shown because they are identical to those of the \MC model.}
\label{fig:autocorr}
\end{figure}

Let $c_i(t)$ denote the state of the $i^{th}$ channel at time $t$, where $c_i(t)=1$ indicates an open channel and  $c_i(t)= 0$ indicates that the channel is closed.  The autocorrelation for the fraction of open channels then becomes
\begin{align}
\label{eq_autocorrsum}
R(t)  &= \frac{\mbox{E}\left[ \frac{1}{N^2}\sum_{i,j=1}^N c_i(t)c_j(0) \right]-\mu_\chan^2}{\sigma_\chan^2}  \\
	&=\frac{\mbox{E}\left[  c_i(t)c_i(0) \right]-\mu_\chan^2}{\sigma_\chan^2}. \notag
\end{align}
This simplification is possible because the \MC model assumes that all channels are statistically identical and independent.
The only unknown term in Eqn.~\ref{eq_autocorrsum} is $\mbox{E}\left[ c_i(t)c_i(0)\right]$.  Since $c_i$ is a binary random variable, expected value of $c_i(t)c_i(0)$ is equal to the probability that the channel is open at the initial time and is also open at the later time $t$.  This probability can be determined by solving the master equation \eqref{eq_mcmaster}, which is possible in voltage clamp because this system of ordinary differential equations is a linear equation with constant coefficients.  The probability that the channel is open is given by $p_4$ in Eqn.~\ref{eq_mcmaster}, so $\mbox{E}\left[ c_i(t)c_i(0)\right]$ is equal to the entry in the last row and the last column of the matrix exponential of the matrix in \eqref{eq_mcmaster}.  We find:
\begin{equation}
\label{eq_mcautocorr}
R(t) = \frac{4 \alpha_n^3e^{-\frac{t}{\tau_n}} + 6 \alpha_n^2 \beta_n e^{-\frac{2t}{\tau_n}} +  4 \alpha_n \beta_n^2 e^{-\frac{3t}{\tau_n}} +   \beta_n^3 e^{-\frac{4t}{\tau_n}}}{(2\alpha_n + \beta_n)(2 \alpha_n^2 + 2 \alpha_n \beta_n +\beta_n^2)},
\end{equation}
where $\tau_n = \frac{1}{\alpha_n+\beta_n}$.  The solid black line in panel (b) of Fig. \ref{fig:autocorr} shows this function for a voltage clamp value of $V=0$~mV.  The same analysis was applied to the \Nap channel, the result of which is also shown in panel (a) of Fig. \ref{fig:autocorr}.  Note the difference in time scales on the ordinate in subpanels (a) and (b). Temporal correlations in the fraction of open \Nap channels decay rapidly within the first millisecond, whereas the temporal correlation in the fraction of open \Kp channels persists for nearly 10~ms.

Equation \ref{eq_mcautocorr} reveals an important feature of the \MC models---the structure of the channel defines the temporal profile of the channel noise statistics.  In the case of the \Kp channel, the transitions between the five possible channel configurations induce correlations on four time scales, the first four multiples of $1/\tau_n$.  The \Nap channel has eight possible channel configurations because it has three $m$ subunits and one $h$ subunit, so the autocorrelation function for the fraction of open \Nap channels has seven time scales.

%
%
%

\subsubsection{Channel SDE Model}
Using the same approximations that allow the channel models to be described as a multivariable OU process, we can compute the autocorrelation function for the proportion of open channels using standard methods \cite{Gardiner2004}.  We find that the autocorrelation function is identical to Eqn.~\ref{eq_mcautocorr} and thus do not include it in Fig. \ref{fig:autocorr}

\subsubsection{Subunit SDE Models}
We compute the autocorrelation for the \SJ and \FLs models using the OU approximation of $n$ in voltage clamp.  We derive our results using the Ito calculus.  The well-known solution to the OU process in Eqn.~\ref{eq_nou} at long times is \cite[e.g]{Gardiner2004}:
\begin{equation}
\label{eq_nsolution}
n(t) =  \mu_\sub + \int_0^t \bar \sigma_n e^{-\frac{t-u}{\tau_n}} dW_u.
\end{equation}
To calculate the autocorrelation in the \SJ model, we take the expectation of the product of four independent solutions of the form of Eqn.~\ref{eq_nsolution} and normalize by the voltage clamp mean and variance shown in Eqns.~\ref{ESJ} and \ref{VSJ}:
\begin{align}
\label{eq_sjautocorr}
R_\SJ(t) 	&= \frac{ \E [ \Pi_{i=1}^4 n_i(t)n_i(0)]  - \E_\SJ(f)^2}{\Var_\SJ(f)}\notag \\
		&=\frac{4 \mu_\sub^6 e^{-\frac{t}{\tau_n}} + 6 \mu_\sub^4 \sigma_\sub^2 e^{-\frac{2t}{\tau_n}} }{4 \mu_\sub^6 + 6 \mu_\sub^4 \sigma_\sub^2 } + \BigOh(\textstyle{\frac{1}{N^2}}).
\end{align}
For brevity, we omit terms of order higher than 1/N.

For the \FLs model, we instead take a single solution of the form of Eqn.~\ref{eq_nsolution} and raise it to the fourth power and normalize by the voltage clamp mean and variance shown in the Eqns.~\ref{EFL} and \ref{VFL}.  The autocorrelation of this function is:
\begin{align}
\label{eq_flautocorr}
R_{\FLs}(t)	&  = \frac{ \E[n(t)^4 n(0)^4] -\E_\FLs(f)^2 }{\Var_\FLs(f)} \notag \\
		& = \frac{ (2 \mu_\sub^6 + 12 \mu_\sub^4 \sigma_\sub^2  )e^{-\frac{t}{\tau_n}} + 9 \mu_\sub^4 \sigma_\sub^2 e^{-\frac{2t}{\tau_n}} }{2 \mu_\sub^6 + 21 \mu_\sub \sigma_\sub^2 } + \BigOh(\textstyle{\frac{1}{N^2}}). 
\end{align}

Calculation of the higher order terms shows that the same exponential time scales (the first four multiples of $1/\tau_n$) are present in $R(t)$ for all of the models, but the coefficients are different.  In particular, Eqns.~\ref{eq_sjautocorr} and \ref{eq_flautocorr} show that, to leading order in $1/N$, the subunit models lack the faster time scales of the \Kp channel.  As a result, temporal correlations in the subunit-based channel noise models persist longer than those of the \MC model and the channel-based SDE model.  
Fig. \ref{fig:autocorr}, voltage clamped at $V=0$~mV, displays these differences in the autocorrelation functions of the subunit models.  The dotted line shows the result for the \SJ model (independent subunit populations) and the dashed line shows the result for the \FLs model (identical subunit populations).  The decay time in the autocorrelation of the proportion of open channels is longer for the subunit models for both the \Nap channel shown in panel in (a) and the \Kp channel shown in panel (b).

\section{Numerical Simulations} \label{sec:Simulations}
In this section we report results from numerical simulations of the \MC model as well as the three SDE models analyzed above.  We first verify the results of our analysis of voltage clamp statistics and then measure the statistics of interspike intervals in order to test how well the SDE models replicate the stochastic features of the \MC model when voltage is allowed to evolve freely according to Eqn.~\ref{eq_HH}.   In all simulations we use the parameter values listed in Table \ref{table_parameter}.  We perform simulations for three different membrane areas: 1, 10, and 100 $\mu m^2$.  The corresponding channel counts are shown in Table \ref{table_area}.

\subsection{Methods} \label{sec:numericalmethod}

\begin{table}
\centering
\caption{Parameter values from \cite{Hodgkin1952}.  Note that resting potential has been shifted to 0 mV.}
\label{table_parameter}
  \begin{tabular}{ l c c }
    \hline
    Symbol & Definition & Value, units \\ \hline \hline
    $C$ & Membrane capacitance & 1 $\mu$F / cm$^2$ \\ 
    $\bar g_\Na$ & Maximal sodium conductance & 120 mS/cm$^2$  \\ 
    $\bar g_\K$ & Maximal potassium conductance  &  36 mS/cm$^2$  \\ 
    $g_{\textrm L}$ & Leak conductance &  0.3 mS/cm$^2$  \\ 
    $E_\Na$ & Sodium reversal potential & 115 mV   \\ 
    $E_\K$ & Potassium reversal potential  & $-12$~mV   \\ 
    $E_{\textrm L}$ & Leak reversal potential  &  10.6 mV   \\ 
    $\rho_\Na$ & Sodium channel density  &  60 / $\mu m^2$  \\ 
    $\rho_{\K}$ &  Potassium channel density & 18 / $\mu m^2$  \\ 
    \hline
  \end{tabular}
\end{table}

\begin{table}
\centering
\caption{Membrane areas and corresponding channel counts used in numerical simulations}
\label{table_area}
  \begin{tabular}{ c | c  c }
    \hline
    Membrane Area\quad & \multicolumn{2}{c}{\# of channels}\\ \cline{2-3}
    ($\mu m^2$) & \quad\Nap\quad & \quad\Kp\quad \\ \hline \hline
    1 	& 60  & 18  \\
    10 	& 600 & 180 \\
    100 & 6000& 1800\\ 
    \hline
  \end{tabular}
\end{table}

Sample Fortran code used to simulate the four stochastic models is available at the ModelDB website (accession number 128502) \cite{Goldwyn2010modeldb}.  All simulations used a time step of size 0.01~ms.  We define a spike using two conditions: the membrane potential must exceed 60 mV and it must have remained below 60~mV for the previous 2~ms (approximately the width of a spike).  To generate Gaussian random numbers, we first produced uniform random numbers using the Mersenne Twister algorithm \cite{mt19937:FortranVer} and then transformed these to Gaussian random numbers using the Box-Muller method \cite{Press1988}.

In Section \ref{sec:isiresults}, we characterize the spiking response of the different models in response to stimuli of the form:
\begin{equation}
\label{eq_I}
I(t) = \Idc + I_{noise}\xi_t, 
\end{equation}
where $\xi_t$ is a Gaussian white noise process with zero mean and unit variance.  This type of input is commonly used to characterize the response of stochastic Hodgkin-Huxley models \cite[e.g.]{Rowat2007,Sengupta2010}.  The additive white noise term can be interpreted as a simplified method for representing the combined effect of numerous synaptic inputs that neurons in cortex and other networks receive \emph{in vivo}, see for instance \cite{Brunel2001}.  
We simulate spike trains for varying membrane area, DC input current $\Idc$, and input noise $I_{noise}$.  We report the mean and coefficient of variation (CV) for the first 2000 interspike intervals (ISI) for each spike train.

\subsubsection{Markov Chain Model} 
The Markov chain describing each \Kp channel is shown in Eqn.~\ref{eq_KMarkovChain}.  The Markov chain that governs the state of each \Nap channel includes three $m$ subunits and one $h$ subunit and is therefore described by an eight state Markov chain:
\begin{equation}
\label{eq_NaMarkovChain}
\begin{array}{ c c c c c c c}
  0,0 &  \overset{3\alpha_m}{\underset{\beta_m}{\rightleftarrows}}  &   1 ,0 &  \overset{2\alpha_m}{\underset{\beta_m}{\rightleftarrows}} &   2 ,0 &  \overset{\alpha_m}{\underset{\beta_m}{\rightleftarrows}} &  3,0 \\
\scriptstyle{\alpha_h} \downarrow \uparrow \scriptstyle{\beta_h}& &  \scriptstyle{\alpha_h} \downarrow \uparrow \scriptstyle{\beta_h} & &   \scriptstyle{\alpha_h} \downarrow \uparrow \scriptstyle{\beta_h}& &    \scriptstyle{\alpha_h} \downarrow \uparrow \scriptstyle{\beta_h} \\
  0 ,1 &  \overset{3\alpha_m}{\underset{\beta_m}{\rightleftarrows}}  &   1 ,1 &  \overset{2\alpha_m}{\underset{\beta_m}{\rightleftarrows}} &   2 ,1 &  \overset{\alpha_m}{\underset{\beta_m}{\rightleftarrows}} &  3 ,1
\end{array}
\end{equation}
The channel is in the conducting state when all three $m$ subunits and the $h$ subunit are open.
The voltage-dependent transition rates for the $m$ and $h$ subunits are \cite{Hodgkin1952}:
\begin{align}
\alpha_m(V) &= 0.1  \frac{25-V}{\exp(\frac{25-V}{10})-1} \notag \\
\beta_m(V) &= 4 \exp(-\frac{V}{18})\notag \\
\alpha_h(V) &=0.07 \exp (-\frac{V}{20})\notag \\
\beta_h(V) &= \frac{1}{  \exp(\frac{30-V}{10}+1) } \notag 
\end{align}

The Markov chains in Eqns.~\ref{eq_KMarkovChain} and \ref{eq_NaMarkovChain} define the possible states of each individual channel.  Rather than simulating individual channels in the membrane patch, however, it is more efficient to track the number of channels in each state using the Gillespie algorithm \cite{Gillespie1977, Chow1996}.  At each time step, the fraction of open \Nap channels $f_\Na$ and \Kp channels $f_\K$ is computed and the voltage is updated using the forward Euler algorithm applied to Eqn.~\ref{eq_HH}.

\subsubsection{Channel SDE Model (\FLs)}
\label{numeric:channel}
The channel SDE model is a system of twelve differential equations derived by Fox and Lu \cite{Fox1994}.  In matrix form, it can be written as:
\begin{align} \label{eq_channelsdenumeric}
C \frac{dV}{dt} &=  -\bar g_{\Na} y_{31} (V-E_{\Na}) - \bar g_\K x_4 (V-E_\K) - g_\textrm{L} (V-E_\textrm{L}) + I  \notag \\
\frac{d \vect{x}}{dt} &=  A_\K \vect{x} + 4 \alpha_n x_0 \vect{e_1} + S_\K \vect{\xi}_\K \\
\frac{d \vect{y}}{dt} &=  A_\Na \vect{y} + 3 \alpha_m y_{00} \vect{e_1}+ \alpha_h y_{00} \vect{e_4}+ S_\Na \vect{\xi}_\Na \notag.
\end{align}
The vector $\vect{x}$ is made up of entries $x_i$ ($i=1,2,3,4$) that represent the proportion of \Kp channels with $i$ open $n$ subunits.  The entries of $\vect{y}$ are denoted $y_{ij}$ ($i=0,1,2,3$ and $j=0,1$) and represent the proportion of \Nap channels with $i$ open $m$ subunits and $j$ open $h$ subunits.  The vectors $\vect{e_i}$ are column vectors with a one in the $i^{th}$ entry and zero elsewhere.    
Following Fox and Lu, we use the fact that $\sum_{i=0}^4 x_i= 1$ and $\sum_{i=0}^3 \sum_{j=0}^1 y_{ij} = 1$ to define $x_0$ and $y_{00}$.  This allows us to reduce the dimension of the system of SDEs from fourteen to twelve.  We note that this reduction of dimension is exact, following from properties of the $A$ and $S$ matrices.
We include the details of the matrices $A_\Na$, $A_\K$, $S_\Na$ and $S_\K$ in the Appendix \ref{appendix:sde}.

The values of the $x_i$ and $y_{ij}$ represent proportions of channels in a particular configurations so, in theory, they should lie within five-dimensional and eight-dimensional hypercubes bounded by the intervals $[0,1]$.  Moreover, since the $x_i$ and $y_{ij}$ each sum to one, the values of these variables should in fact lie on hyperplanes within these hypercubes.  If, in the course of numerical simulations, there are excursions of $\vect{x}$ and $\vect{y}$ off of these high dimensional bounded surfaces, then the solution will lack biological meaning because a value that represents the proportions of channels should not be negative or exceed one.  A numerical difficulty also arises when these variables do not lie on the proper bounded hyperplanes.   To define $S_\K$ and $S_\Na$, one needs to take matrix square roots of diffusion matrices in every time step.  If the $\vect{x}$ and $\vect{y}$ do not lie on the bounded surfaces, then the diffusion matrices, which depend on the values of $\vect{x}$ and $\vect{y}$, will no longer be guaranteed to be positive semi-definite, which may make it impossible to compute real valued matrix square roots.

In principle, it may be possible to incorporate a projection or a reflection into the numerical method to ensure that $\vect{x}$ and $\vect{y}$ remain on these bounded, high-dimensional surfaces of admissible values.  We demonstrate that a simpler approach in which the individual values of the $x_i$ and $y_{ij}$ are not confined within $[0,1]$, but rather are free to evolve without boundary conditions, gives an adequate numerical approximation to the interspike interval statistics of the \MC model.  With this simplification, there is no longer a guarantee that real matrix square roots of the diffusion matrices will exist, so we replace the values of $x_i$ and $y_{ij}$ in the diffusion matrices with their equilibrium values.  The validity of approximation is discussed in Appendix \ref{appendix:noise}.   After implementing the above simplifications, we solved the resulting system of SDEs using the Euler-Maruyama method \cite{Higham2001}.  Comparing our results with implementations that bound the SDE solutions would be an interesting subject for future work, but is beyond the aims of this paper.

\subsubsection{Subunit SDE Models}
\label{numeric:subunit}
The two subunit SDE models that we study are the independent subunit (\SJ) model:
\begin{align}
C \frac{dV}{dt} &= -\bar g_\Na m_1 m_2 m_3 h (V-E_\Na) \notag \\
&- \bar g_\K n_1 n_2 n_3 n_4(V-E_\K) -g_\textrm{L} (V-E_\textrm{L}) + I  \notag \\
\frac{dm_i}{dt} &= \alpha_m (1-m_i) - \beta_m m_i + \sigma_m(V) \xi_{m_i}(t), \mbox{ where }i=1,2,3 \notag \\
\frac{dh}{dt} &= \alpha_h (1-h) - \beta_h h + \sigma_h(V) \xi_h(t) \notag \\
\frac{dn_i}{dt} &= \alpha_n (1-n_i) - \beta_n n_i + \sigma_n(V) \xi_{n_i}(t), \mbox{ where }i=1,2,3,4\notag .
\end{align}
and the identical subunit (\FLs) model
\begin{align}
C \frac{dV}{dt} &= -\bar g_\Na m^3 h (V-E_\Na) - \bar g_\K n^4(V-E_\K) -g_\textrm{L} (V-E_\textrm{L}) + I  \notag \\
\frac{dm}{dt} &= \alpha_m (1-m) - \beta_m m + \sigma_m(V) \xi_{m}(t) \notag\\
\frac{dh}{dt} &= \alpha_h (1-h) - \beta_h h + \sigma_h(V) \xi_h(t) \notag \\
\frac{dn}{dt} &= \alpha_n (1-n) - \beta_n n + \sigma_n(V) \xi_{n}(t) \notag.
\end{align}
The difference between these two models is that, in the former, we compute multiple independent realizations of the $n$ and $m$ type subunits and the product of these terms enter into the equation for $V$ whereas in the latter, the approximation is made that all subunit classes are assumed to be perfectly correlated so only one SDE is solved for each subunit type and the solution is raised to the appropriate power (4 for $n$ and 3 for $m$).
The gating variables represent proportions of open subunits so we enforce boundary conditions that prevent the values of the gating variables from exceeding one or becoming negative.

We note that the form of $\sigma_x^2$ ($x=m,h,\mbox{ or }n$) that we use is given in Eqn.~\ref{eq_sigman}.  In particular, the noise terms depend on voltage as well as the subunit variables themselves.  We do not apply the equilibrium noise approximation in our simulations of the subunit SDEs, although this approximation has been used in past simulation studies \cite{Fox1994, Bruce2009, Sengupta2010}.
We solve these systems of SDEs using the Euler-Maruyama method \cite{Higham2001}.

\subsection{Simulation Results: Voltage Clamp} \label{sec:vclampresults}
In Figs.~\ref{fig:vclamp_mean_std} and \ref{fig:autocorr}, we compare results from numerical simulations for a membrane patch size of 10 $\mu m^2$ against the analytical calculations presented in Section \ref{sec:vclamp}.  To simulate the voltage clamp condition, we fix $V$ at a particular value and keep it constant throughout the simulation.  Figure \ref{fig:vclamp_mean_std} shows the mean and standard deviation of the proportion of open \Nap and \Kp channels as a function of the voltage clamp value.  In most cases, the values computed from numerical simulations (symbols) match the analytical results (lines).  Of particular note is the fact that the computed values for the \MC 
model (circle) and the \FLc model (x) are virtually indistinguishable.

The only deviation between the numerical results and the analytical solutions occurs in the subunit models for the mean values of the \Nap channels at high voltage values.  The cause of this discrepancy is that the analytical treatment assumes that $n$ is Gaussian-distributed whereas in the numerical methods the values of $n$ are bounded between zero and one.  For high voltage values of $V$, the proportion of open \Nap channels is very small and the variance is non-zero so approximating the distribution of $n$ as a Gaussian will allow $n$ to take negative values.  This cannot occur in the numerical simulations so the theoretical value for the mean fraction of open \Nap channels will be less than the simulated value.  As the number of channels increase, the variance of the fraction of open \Nap channels decreases, which decreases the probability that a Gaussian-distributed $n$ will take negative values.  The discrepancy between the analytically and numerically calculated values for the mean fraction of open \Nap channels will decreases, therefore, as the number of \Nap channels increases.

\subsection{Simulation Results: Interspike Intervals} \label{sec:isiresults}

\begin{figure}[!tbp]
\includegraphics[width=8.6cm]{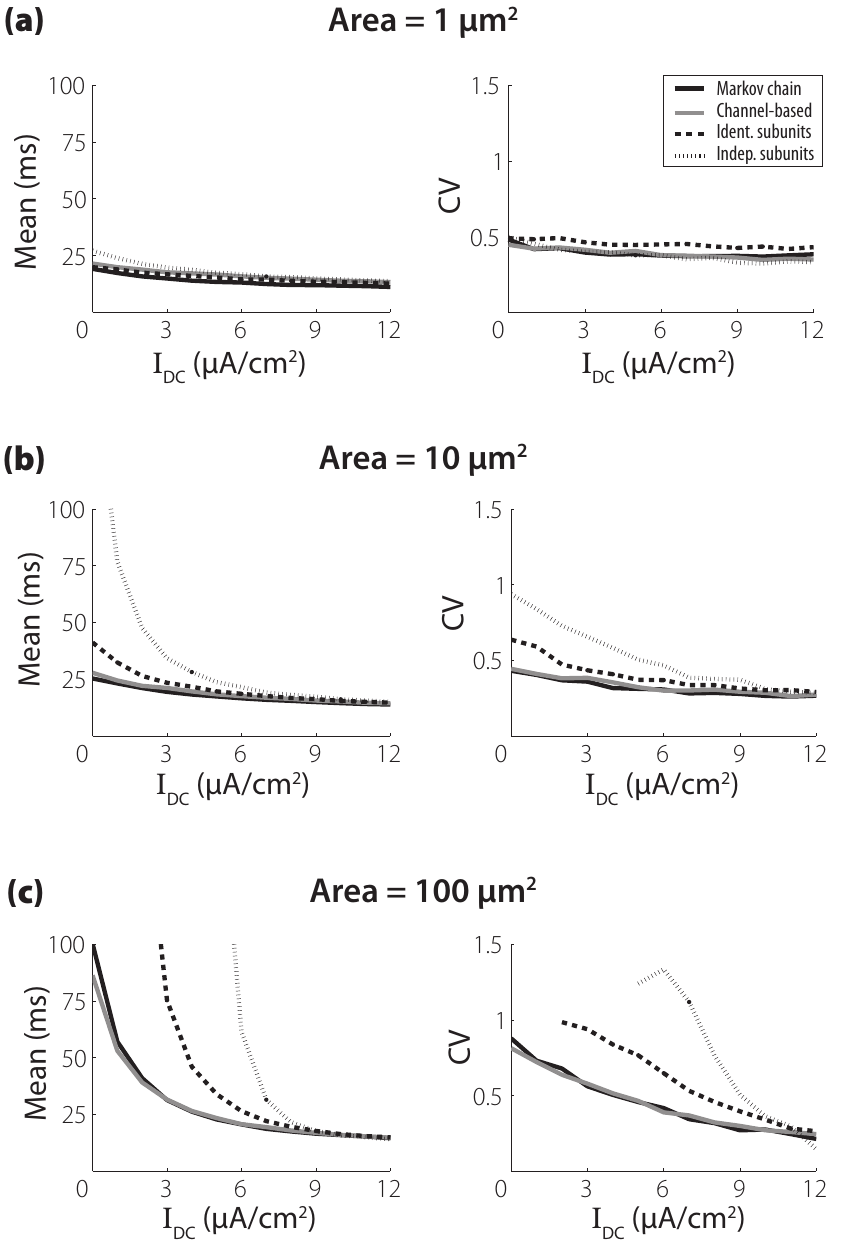}
\caption{Means and coefficients of variation (CV) of first 2000 interspike intervals as a function of the DC current level (abscissa) for a constant current input.  \textbf{(a)} Results for a membrane area of 1~$\mu m^2$.  \textbf{(b)} Results for a membrane area of 10~$\mu m^2$.  \textbf{(c)} Results for a membrane area of 100~$\mu m^2$.}
\label{fig:isistat}
\end{figure}

Figure \ref{fig:isistat} shows mean and CV (left and right columns, respectively) of ISIs for three membrane areas that increase from top to bottom in each column.  The input to the model is a constant DC input ($I_{noise}=0$ in Eqn.~\ref{eq_I}).
The value of \Idc is shown on the x-axis.  In general, these simulations show that the rate and regularity of spiking activity produced by the \MC (black line) and \FLc (gray line) models are in close quantitative agreement whereas the \FLs (dashed line) and \SJ (dotted line) models produce, on average, dramatically longer ISIs.  The behavior of the models is most disparate for low stimulus levels and larger membrane areas.  In these cases, the stimulus is not sufficiently strong to drive regular firing, so spike events are predominantly determined by stochastic fluctuations in the conductances due to channel noise.  As the DC current is increased, the models respond to the external stimulus and there is a smaller effect of channel noise on spike timing.  This leads all models to exhibit similar to mean ISI and CV values at high current levels.

The mean ISIs for the subunit SDE models exceed those of the \MC model for all stimulus conditions and membrane areas.  This has been previously observed for the \FLs model \cite{Sengupta2010}.  In the voltage clamp analysis we found that the variance in the proportion of open \Kp and \Nap channels is much smaller for both of these models than for the \MC model.  This lack of conductance fluctuations leads to reduced spike rates at low stimulus levels, relative to the \MC model.

Overall, the results for the \FLc model show that it is possible to approximate the \MC model with SDE and still obtain quantitatively accurate results, even for small numbers of channels.  This is an important and a nontrivial result---the system size expansion is formally valid only in the limit of infinitely many channels, but we show here that it can be applied to a small number of channels, where membrane fluctuations have a major impact on spiking statistics.  Moreover, the equilibrium noise approximation and treatment of boundary conditions do not appear to substantially degrade solution accuracy over a wide range of stimuli.

Nevertheless, there are some discrepancies between the \FLc and \MC models.  At the smaller membrane areas ($1~\mu m^2$ and $10~\mu m^2$), the mean ISIs for the \FLc model tend to be longer than the \MC model ISIs.  At the largest membrane area tested, for the case of weak or no input current, this trend is reversed and the \FLc model has shorter mean ISIs than the \MC model.  A possible source for these between the \FLc and \MC models is our treatment of the boundary conditions in the \FLc model and the equilibrium noise approximation. Further investigation of this approximation is needed, but the similar ISI statistics obtained with the \FLc and \MC models suggest that this approximation may be suitable in many cases.

\begin{figure}[!tbp]
\includegraphics[width=8.6cm]{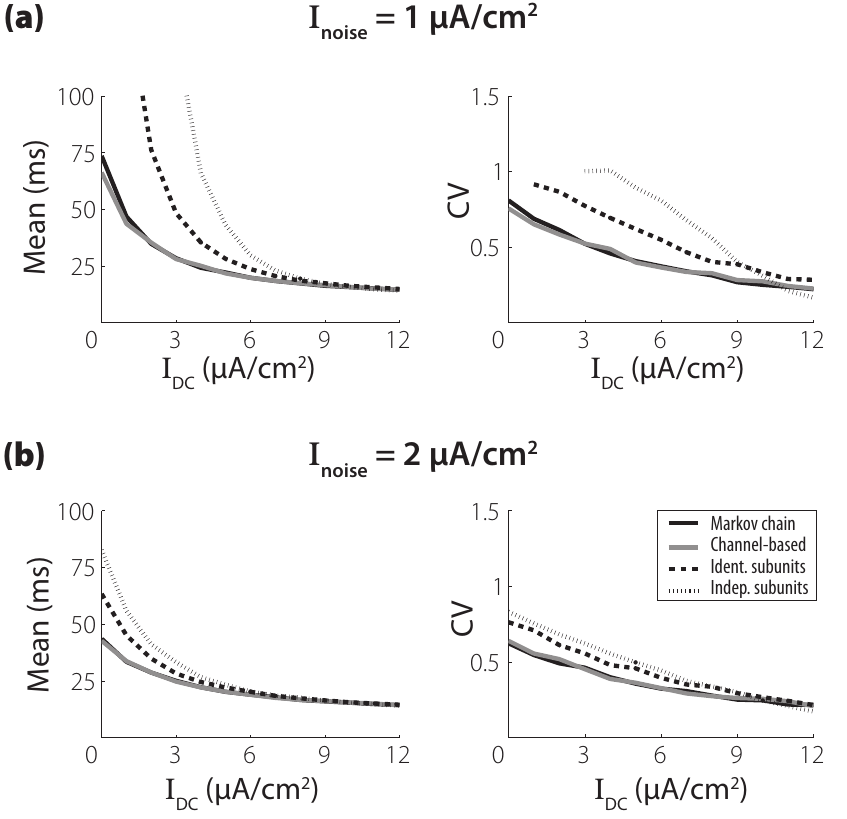}
\caption{Means and coefficients of variation (CV) of first 2000 interspike intervals as a function of the DC current level (abscissa) for a current input of the form $\Idc + I_{noise} \xi_t$ where $\xi_t$ is a Gaussian white noise process with mean zero and unit variance.  Membrane area is 100~$\mu m^2$. \textbf{(a)} and \textbf{(b)} show results for $I_{noise}=1$ and $2~\mu m^2$, respectively.}
\label{fig:isistatnoise}
\end{figure}

Figure \ref{fig:isistatnoise} shows results obtained for two different levels of input noise added to the DC current amounts shown on the x-axis.  We only present ISI statistics for the membrane size of $100~\mu m^2$ because the smaller membrane areas produce the same qualitative differences among the models.  Overall, the effect of the stimulus noise is to reduce the mean ISIs.  Importantly, the ISI statistics of the responses of the \MC and \FLc SDE models to these noisy stimuli remain quantitatively similar.  In fact, the stimulus fluctuations elicit spikes even at low or no DC current levels, so the differences in the mean ISIs between the \MC and \FLc SDE models become less apparent.  Overall, this result indicates that the equilibrium noise approximation does not break down in the presence of a rapidly fluctuating external stimulus.  Finally, the results for the subunit SDEs show, once again, that the stochastic dynamics and spiking activity of these approximate models do not accurately replicate the statistics of the \MC model.

%
%
%

%
%
%

\section{Discussion} \label{sec:discussion}

Beginning with the work of Fox and Lu \cite{Fox1994, Fox1997}, the question of whether SDE models of channel noise can accurately approximate \MC models has been explored.  SDE models of membrane voltage fluctuations in HH models have several attractive features, including possible improvements in the speed of numerical simulations and the opportunity to analyze these models using nonlinear SDE theory \cite[e.g.]{Freidlin1998}.
In recent years, however, the SDE approach has come under increasing scrutiny.  Numerical simulations of the most commonly used SDE model, which we have called the \emph{identical subunit} model, have shown that this approach produces weaker conductance and voltage fluctuations than the corresponding \MC model \cite{Bruce2009, Faisal2010}.  As a consequence, the firing rates of this SDE model are substantially lower \cite{Sengupta2010} (and, equivalently, the mean ISIs are longer \cite{Zeng2004}), there is less variability in the occurrences and timing of spikes in response to a brief pulse of current \cite{Mino2002}, and information is transmitted at a higher rate \cite{Sengupta2010}.  Furthermore, these discrepancies persist even as the number of channels increases \cite{Zeng2004, Bruce2009, Sengupta2010}.  In sum, there is an emerging consensus in the literature that the \MC model cannot be approximated accurately using a subunit system of SDEs.

We have demonstrated in this paper that an alternative SDE approach that is based on the multi-state structure of each ion channel can approximate the channel noise effects that are present in the \MC model, even for relatively small numbers of channels, as long as the system-size expansion that is used to derive the SDE model is carried out properly.  If one first defines the structure of a channel and then defines the dynamics of the proportions of channels in each configuration, one arrives at the channel-based SDE model (see discussion in Section \ref{subsec:intuitive} and equations in Section \ref{numeric:channel}).  If, instead, one approximates the proportion of subunits in the open or closed states with an SDE, then one obtains a subunit-based SDE model (see discussion in Section \ref{subsec:intuitive} and equations in Section \ref{numeric:subunit}).  Through our analysis of the stationary statistics of the proportion of open channels in voltage clamp, we have shown that the former approach, which we call \emph{channel-based}, can provide a quantitatively accurate approximation to the \MC model.  We have also confirmed that the latter, \emph{subunit-based} approach, should not be considered an approximation of the \MC model because it has fundamentally different stochastic properties from the \MC model.  We conclude that the SDE approach is a valid approximate model for channel noise, but that is necessary to properly define the system of SDEs based on the structure of each channel.  In particular, one cannot include noise in the subunit equations in the manner suggested by Fox and Lu and expect results that are consistent with the \MC ion channel model.

To our knowledge, we are the first to present the numerical results of the channel SDE model \cite{Fox:PersonalComm}.  We simulated three membrane areas (1, 10, and 100 $\mu m^2$ as in \cite{Sengupta2010}), where the number of \Nap channels range from 60 to 6000 and the number of \Kp channels range from 18 to 1800.  Our simulation results show that, in most cases, the ISI statistics for this model in response to constant and noisy current inputs are in close quantitative agreement with the \MC model (see Figs.~\ref{fig:isistat} and \ref{fig:isistatnoise}).  This finding is encouraging because the channel SDE model was derived by Fox and Lu using a system-size expansion \cite{Fox1994,Fox1997} that is formally valid only in the limit of a large number of channels.  We did not find any evidence that the approximation was invalid for the finite populations of ion channels at hand, and note that in many applications, the channel counts are in fact much higher.  For instance, Rowat suggested that typical channel numbers in spike initiation zones may be on the order of $10^4$ to $10^6$ \cite{Rowat2007}.   The channel counts used in the present study may be relevant to applications in which small patches of neural membrane can drive spiking activity.  For example, a node of Ranvier of the auditory nerve fiber can produce a spike in response to cochlear implant stimulation.  Typically, the nodes have surface areas of a few square micrometers \cite{Imennov2009} and are usually modeled with 1000 or fewer \Nap channels \cite{Negm2008, Imennov2009, Woo2010}.

As first pointed out by Fox and Lu \cite{Fox1994} and as discussed in Section \ref{numeric:channel}, numerical simulations of the channel SDE model can be computationally expensive.  One particularly compute-intensive part of the algorithm is calculating matrix square roots to determine stochastic terms in the SDE at every time step of the simulation.  We have performed this operation using the optimized CBLAS library, yet the channel SDE model still required approximately 25 times as much computational time as the \FLs model.  Fortunately, as is the case with all three SDE approaches, the channel model has one considerable advantage over \MC---its computation time does not depend on the number channels.  For example, in our implementation of the Gillespie method, the computational time increased 12-fold as we increased the number of channels from 600~\Nap and 180~\Kp to 6000~\Nap and 1800~\Kp.  We found that even the channel model, slowest of the SDE approaches discussed in this paper, is faster than the \MC model once the number of channels is greater than approximately 1200 \Nap and 360 \Kp.
Furthermore, the computational burden of solving the channel SDE model may be reduced by considering other methods for computing matrix square roots \cite[e.g.]{Allen2007}.  Higher-order SDE solvers than the Euler-Maruyama method could also speed up SDE simulations \cite[e.g.]{Alzubaidi2010}.  

\begin{figure}[!tbp]
\includegraphics[width=8.6cm]{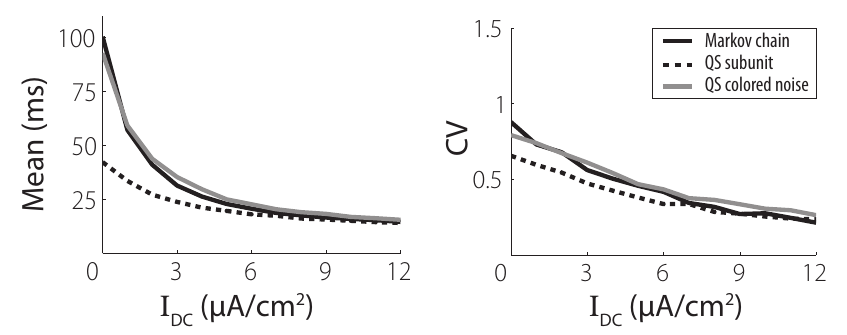}
\caption{Means and coefficients of variation (CV) of first 2000 interspike intervals for the \MC model (black line) and two quasistationary models (see text and Appendix \ref{appendix:qs} for details) in response to constant current input.  DC current level is given by the abscissa.  Membrane area is 100 $\mu m^2$.}
\label{fig:qsmodel}
\end{figure}

Even with increasingly efficient numerical methods for the channel SDE, a stochastic model that more directly resembles the classical HH equations, as opposed to the twelve-dimensional system of SDEs that defines the channel model, is still desirable, as it would connect more directly with a wealth of studies of the original four-dimensional HH equations.  In a simulation study of the Fox and Lu model, Bruce \cite{Bruce2009} sought to derive noise terms numerically for the subunit equations that would ``correct" the Fox and Lu model so that the fluctuations in the proportions of open channels would match the \MC model.  We have performed a similar analysis using the analytical voltage clamp results (see Appendix \ref{appendix:qs} for details).  We can redefine the magnitude of the noise in the subunit equations (\ie $\sigma_n$ in Eqn. \ref{eq_sdesubunit}) to produce a modified subunit SDE model with the same means and variances for the proportion of open channels in voltage clamp, to $\BigOh(1/N)$.  This model is constructed based on our analytical results for the stationary distributions of the proportions of open channels, so we call it a \emph{quasistationary} model.
 The ISI statistics for this modified subunit SDE model are shown by the dashed line in Fig. \ref{fig:qsmodel}.  This demonstrates that adjusting the noise terms in a subunit SDE model in order to fit the variances of the open channels in voltage clamp is not sufficient to provide an improved fit to the spiking dynamics of the \MC model.  

The reason for this can be seen in the multiple time scales of the autocorrelation functions for the \MC model (Eqn.~\ref{eq_mcautocorr}) and subunit models (Eqns.~\ref{eq_sjautocorr} and \ref{eq_flautocorr}).  We can alter the noise terms in the subunit model so that it produces the correct stationary variances of the fractions of open channels, but we cannot modify the autocorrelation functions in a way that makes them consistent with the \MC model.  We therefore formulated a second quasistationary model by adding colored Gaussian noise to the conductances in the HH equations (details of this model are in Appendix \ref{appendix:qs}).  This second quasistationary model produces the stationary distribution of the channel SDE in voltage clamp, so the proportion of open channels in voltage clamp for this model has the same mean, variance, and autocorrelation as the \MC and channel SDE models.  

ISI statistics for this model are shown by the gray line in Fig. \ref{fig:qsmodel}.  We found that this model reproduced the statistics of the \MC model much better than any of the subunit SDE models, so we conclude that temporal correlations in the channel noise play a critical role in influencing spike timing.  Moreover, since the structure of the ion channel determines the history-dependence of the channel noise, a valid channel noise model must properly describe the dynamics of the entire channel and not only the kinetics of individual subunits.   Using numerical simulations, Bruce has pointed out that the subunit model does accurately approximate the \MC model for the case of channels with a single subunit \cite{Bruce2009}.  Our analysis explains this observation because, for channels with one subunit, the channel-based and subunit-based SDE models are mathematically identical.

We close by emphasizing that the models described in this paper do not represent complete descriptions of channel noise.  As is always the case, when one attempts to formulate a mathematical descriptions of complex biological processes, numerous assumptions and simplifications are at play.
As our understanding of the structure and dynamics of these membranes improves, it may be necessary to update and improve our mathematical models of channel noise.  Nonetheless, the central theme of this work will remain relevant: the statistics of channel noise are shaped by the activity of individual ion channels, and therefore the approximation methods must also include information about the states of the channels in order to correctly capture the state and history dependence of channel noise.


%
%
%

\section*{Acknowledgments}
We thank Jay Rubinstein for drawing our interest to this problem and for helpful discussion.  We also thank Jay Rubinstein and Adrienne Fairhall for helpful discussions.
This research has been supported by the National Institute on Deafness and Other Communication Disorders  (F31 DC010306---J.H.G., T32 DC005361---N.S.I.), a fellowship from the Advanced Bionics Corporation (N.S.I.), the McKnight Endowment Fund for Neuroscience (M.F.), and the Burroughs-Wellcome Fund (Career Award at the Scientific Interface---E.S.-B.). 

\cleardoublepage
\appendix

\section{Matrices used in numerical simulations of the channel SDE model}
\label{appendix:sde}
The state vectors are defined as $\vect{x} = [x_1, x_2, x_3, x_4]^T$ and $\vect{y} = [y_{10}, y_{20}, y_{30}, y_{01}, y_{11}, y_{21}, y_{31}]^T$. 
The number of \Kp and \Nap channels are denoted by $N_\K$ and $N_\Na$, respectively.
The matrices $A_\K$ and $A_\Na$ in Eqn.~\ref{eq_channelsdenumeric} are:
\begin{equation} 
A_\K = \left[ \begin{smallmatrix}
-(3\alpha_n + \beta_n)& 2\beta_n &0 &0\\
 3 \alpha_n& -2(\alpha_n + \beta_n) & 3\beta_n  & 0\\
0 & 2\alpha_n &-(\alpha_n + 3\beta_n) & 4\beta_n \\
0 & 0& \alpha_n & -4\beta_n  \end{smallmatrix} \right] . \notag
\end{equation}
\begin{widetext}
\begin{equation}
A_\Na = \left[ \begin{smallmatrix}
-(2*\alpha_m +\beta_m + \alpha_h) &2 \beta_m & 0 &0 &\beta_h &0 &0  \\
 2 \alpha_m& -(\alpha_m + 2\beta_m + \alpha_h) & 3\beta_m &0&0&\beta_h&0\\
0 & \alpha_m &-(3 \beta_m + \alpha_h)  & 0 &0&0&\beta_h\\
0 & 0 & 0 &-(3 \alpha_m + \beta_h ) & \beta_m & 0 &0\\
\alpha_h & 0 &0 & 3 \alpha_m & -(2 \alpha_m + \beta_m + \beta_h) & 2 \beta_m & 0 \\
 0 & \alpha_h &0 & 0 &2 \alpha_m &-(\alpha_m + 2 \beta_m + \beta_h)&3 \beta_m\\
  0 & 0&\alpha_h & 0  &0&\alpha_m& -(3 \beta_m + \beta_h) \end{smallmatrix} \right]  \notag
\end{equation}
\end{widetext}

The matrices $S_\K$ and $S_\Na$ are the square root matrices of the following diffusion matrices:
\begin{widetext}
\begin{equation} 
D_\K = \frac{1}{N_\K} \left[ \begin{smallmatrix}
4 \alpha_n \bar{x}_0 + (3 \alpha_n + \beta_n)\bar{x}_1 + 2 \beta_n \bar{x}_2& -(3 \alpha_n \bar{x}_1 +2 \beta_n \bar{x}_2) & 0& 0\\
-(3 \alpha_n \bar{x}_1 +2 \beta_n \bar{x}_2)  & 3 \alpha_n \bar{x}_1 + 2(\alpha_n + \beta_n) \bar{x}_2 + 3\beta_n \bar{x}_3 & -(2 \alpha_n \bar{x}_2 + 3 \beta_n \bar{x}_3) &0  \\
0 & -(2 \alpha_n \bar{x}_2 + 3 \beta_n \bar{x}_3) & 2 \alpha_n \bar{x}_2 + (\alpha_n + 3\beta_n) \bar{x}_3 + 4 \beta_n \bar{x}_4 & -(\alpha_n \bar{x}_3 + 4 \beta_n \bar{x}_4)\\
0 & 0 & -(\alpha_n \bar{x}_3 + 4 \beta_n \bar{x}_4) & \alpha_n \bar{x}_3 + 4 \beta_n \bar{x}_4
 \end{smallmatrix} \right]  \notag
\end{equation}
\end{widetext}
\begin{widetext}
\begin{align} 
\scriptscriptstyle D_\Na = \frac{1}{N_\Na}\left[ \begin{smallmatrix}
d_1 & -2 (\alpha_m \bar{y}_{10} +\beta_m \bar{y}_{20} )& 0 & 0 & -(\alpha_h \bar{y}_{10} + \beta_h \bar{y}_{11}) & 0 & 0\\
-2 (\alpha_m \bar{y}_{10} +\beta_m \bar{y}_{20} ) & d_2  & -(\alpha_m \bar{y}_{20} + 3 \beta_m \bar{y}_{30}) &0&0& -(\alpha_h \bar{y}_{20} + \beta_h \bar{y}_{21})&0\\
 0 & -(\alpha_m \bar{y}_{20} + 3 \beta_m \bar{y}_{30}) & d_3  & 0 &0&0& -(\alpha_h \bar{y}_{30} + \beta_h \bar{y}_{31})\\
 0 & 0 & 0 &d_4 & -(3 \alpha_m \bar{y}_{01} + \beta_m \bar{y}_{11})& 0 &0\\
-(\alpha_h \bar{y}_{10} + \beta_h \bar{y}_{11})& 0 &0 &  -(3 \alpha_m \bar{y}_{01} + \beta_m \bar{y}_{11}) & d_5 & -2(\alpha_m \bar{y}_{11} + \beta_m \bar{y}_{21}) & 0 \\
 0 &  -(\alpha_h \bar{y}_{20} + \beta_h \bar{y}_{21})&0 & 0 &-2(\alpha_m \bar{y}_{11} + \beta_m \bar{y}_{21})  &d_6 & -(\alpha_m \bar{y}_{21}+ 3\beta_m \bar{y}_{31})\\
 0 & 0&  -(\alpha_h \bar{y}_{30} + \beta_h \bar{y}_{31}) & 0  &0 &-(\alpha_m \bar{y}_{21}+ 3\beta_m \bar{y}_{31}) & d_7
 \end{smallmatrix} \right]  \notag
\end{align}
\end{widetext}
where the elements on the diagonal are:
\begin{align}
d_1 &= 3 \alpha_m \bar{y}_{00} + (2 \alpha_m + \beta_m + \alpha_h) \bar{y}_{10} + 2 \beta_m \bar{y}_{20} + \beta_h \bar{y}_{11}  \notag \\
d_2 &= 2 \alpha_m \bar{y}_{10} + (\alpha_m + 2\beta_m + \alpha_h) \bar{y}_{20}   + 3 \beta_m \bar{y}_{30} + \beta_h \bar{y}_{21}\notag \\
d_3 &= \alpha_m \bar{y}_{20} + (3 \beta_m + \alpha_h) \bar{y}_{30}  + \beta_h \bar{y}_{31} \notag \\
d_4 &= \alpha_h \bar{y}_{00} + (3 \alpha_m + \beta_h) \bar{y}_{01} + \beta_m \bar{y}_{11}\notag \\
d_5 &= \alpha_h \bar{y}_{10} + 3 \alpha_m \bar{y}_{01} + (2 \alpha_m + \beta_m + \beta_h) \bar{y}_{11}  + 2 \beta_m \bar{y}_{21} \notag \\
d_6 &= \alpha_h \bar{y}_{20} + 2 \alpha_m \bar{y}_{11} + (\alpha_m + 2 \beta_m + \beta_h)\bar{y}_{21} + 3 \beta_m \bar{y}_{31}\notag \\
d_7 &= \alpha_h  \bar{y}_{30} + \alpha_m \bar{y}_{21} + (3 \beta_m + \beta_h) \bar{y}_{31} \notag 
\end{align}
As discussed, we use the equilibrium mean values of $\vect{x}$ and $\vect{y}$ in the diffusion matrices.  They are:
\begin{equation}
\bar{x}_i =  {4 \choose  i} \frac{\alpha_n^i \beta_n^{4-i}}{(\alpha_n + \beta_n)^4} \notag
\end{equation}
and
\begin{equation}
\bar{y}_{ij} =  {3 \choose  i} \frac{\alpha_m^i \beta_m^{3-i} \alpha_h^j \beta_h^{1-j}}{(\alpha_m + \beta_m)^3(\alpha_h + \beta_h)}. \notag
\end{equation}

\section{Equilibrium Noise Approximation}
\label{appendix:noise}

\subsubsection{Voltage Clamp}The equilibrium noise approximation in voltage clamp can be justified using a small noise expansion \cite{Gardiner2004}.
Let us define the small parameter $\epsilon = 1/N$ and fix the membrane potential at a voltage clamp value so that $\alpha_n$ and $\beta_n$ can be treated as constants.
Assume that $n$ can be written as a series in the small noise parameter $\epsilon$:
\begin{equation}
n(t) = n_0(t) + \epsilon n_1(t) +  \epsilon^2 n_2(t)\cdots. \notag
\end{equation}
If we plug the small noise expansion into the subunit SDE for $n$ in Eqn.~\ref{eq_sdesubunit} and collect terms of $\BigOh(1)$, we find:
\begin{equation}
\label{eq_dn0dt}
\frac{dn_0}{dt} = \alpha_n(1-n_0)-\beta_n n_0, 
\end{equation}
and for terms of $\BigOh(\epsilon)$:
\begin{equation}
\label{eq_dn1dt}
\frac{d n_1}{dt}  = - \alpha_n n_1 - \beta_n n_1 + \sqrt{\alpha_n (1-n_0) + \beta_n n_0}\,\xi(t). 
\end{equation}

Equation \ref{eq_dn0dt} shows that $n_0$ satisfies a deterministic equation that does not depend on stochastic fluctuations in $n$.  In the context of analyzing the stationary distribution of $n$, therefore, we are justified in replacing $n_0$ with its equilibrium value $\frac{\alpha_n}{\alpha_n + \beta_n}$.  If we make this substitution for $n_0$ in the equation for $n_1$ (Eqn.~\ref{eq_dn1dt}) and then form the sum $n_0 + \epsilon n_1$, it is straightforward to arrive at the OU process in Eqn.~\ref{eq_nou}.  The equilibrium noise approximation is therefore the $\BigOh(1/N)$ approximation of the long-time behavior of $n(t)$ in voltage clamp.
The same argument applies for the $m$ and $h$ subunits as well as for the multivariate SDE that defines the channel SDE model.

\subsubsection{Time-dependent voltage}
Fox and Lu suggested applying this approximation in all cases, not just voltage clamp \cite{Fox1994, Fox1997} and we have used the approximation to simplify the numerical methods for solving the channel SDE model.  When $V$ is not in voltage clamp, it evolves naturally and complicates the small noise expansion because it can introduce additional stochastic fluctuations into the gating variables and the voltage-dependent functions transition rate functions. Fox argued that the approximation would be accurate if the relaxation of $V$ to its equilibrium value occurred on a much slower time scale than the relaxation of the gating variables (for the case of the subunit SDE model) \cite{Fox1997}.  Unfortunately, this separation of time scales does not appear to be a generic feature of HH models.
Nevertheless, as shown in Figs.~\ref{fig:isistat} and \ref{fig:isistatnoise}, the equilibrium noise approximation appears to be sufficiently accurate to reproduce spiking statistics to a high degree of accuracy.

\section{Quasistationary Models} \label{appendix:qs}

In the Discussion (Section \ref{sec:discussion}), we introduced two models that we discuss in greater detail here.  We refer to both models as \emph{quasistationary} approximations because they rely on results from our analysis of the stationary statistics of open channels in Section \ref{sec:vclamp}.  In the first model, we were motivated by \cite{Bruce2009} to attempt to improve the accuracy of the subunit SDE model by modifying the noise terms in the gating equations.  We have shown that the stationary variances for the proportion of open channels in the subunit SDE models does not match those of the \MC model.  To correct for this discrepancy, we can redefine $\sigma_n^2(V)$ in Eqn.~\ref{eq_sdesubunit} to guarantee that the stationary variance of $n^4$ matches the stationary variance of the proportion of open \Kp channels under the \MC models.
The problem is simplified if we invoke the equilibrium noise approximation and use the facts that $\Var[n^4] = \E[n^8]-\E[n^4]^2$ and that these higher moments of the stationary distribution of $n$ are known since in voltage clamp $n$ is an OU (Gaussian) process with mean $\mu_\sub = \frac{\alpha_n}{\alpha_n + \beta_n}$ and variance $\sigma^2_n(V) \tau_n(V) /2$.  The final step is to set $\Var[n^4] = \Var[f_\chan]$ and solve for $\sigma_n(V)$.  The exact solution would require inverting a nonlinear equation, but if we neglect terms that are of higher order than $\BigOh(1/N^2)$, we find $\sigma_n(V)$ by finding the roots of $168 \mu_n^4 \sigma_n^4+ 16 \mu_n^6 \sigma_n^2 -(\mu_n^4 (1-\mu_n^4)) / N$, which is quadratic in $\sigma_n^2$.  The same approach was also used to derive a new expression for $\sigma_m$ so that this model also had the same stationary variance for the \Nap channel.  The formula for $\sigma_h$ was left unchanged.

As explained in the Discussion, this first approach did not provide a satisfactory approximation to the \MC model, so we formulated a second quasistationary approximation.  We constructed this second model so that it and the \MC model would have the same autocorrelation functions for the proportion of open channels and the same means and variances in voltage clamp.  The mathematical structure of this model is somewhat unusual in that the conductance is defined as the sum of a deterministic part (given by the HH equations for $m$, $h$, and $n$) and a colored Gaussian processes that is defined by the autocovariance function for the proportion of open channels in the \MC model.  As usual, we illustrate our approach with the \Kp channel.  The conductance is defined to be:
\begin{equation}
\label{eq_gkQS}
g_\K = \bar g_\K (n^4 + \sigma_\chan \eta_t)
\end{equation}
where $n$ is the classical (deterministic) gating variable satisfying an ordinary differential equation of the form of Eqn.~\ref{eq_submaster} and $\eta_t$ is a stochastic process.  To define $\eta_t$, first note that the channel SDE model provided a quantitatively accurate approximation of the \MC model, so it is reasonable to describe the stationary conductance as a Gaussian process.  Second, recall from Eqn.~\ref{eq_mcautocorr} that the autocorrelation function for the proportion of open \Kp channels has four distinct time scales (the first four multiples of $1/\tau_n$).  Taken together, these facts lead us to model the \Kp conductance as a non-Markovian Gaussian process.  The representation theory of Gaussian processes furnishes a systematic method for constructing the stochastic process $\eta_t$ based on the autocorrelation function for the \Kp channel \cite{Hida1993}.  In particular, this theory guarantees that $\eta_t$ can be written in terms of a stochastic (Wiener) integral of the form:
\begin{equation}
\eta_t = \int_0^t \sum_{i=1}^4 a_i(V) e^{-\frac{i(t-u)}{\tau_n(V)}} dW_u \notag
\end{equation}
The coefficients $a_i$ are voltage-dependent and are computed by solving the system of nonlinear equations:
\begin{align}
\tau_n(V) a_1(V) \sum_{i=1}^4 \frac{a_i(V)}{i+1} &= 4 \alpha_n(V)^3 \Gamma \notag \\
\tau_n(V) a_2(V) \sum_{i=1}^4 \frac{a_i(V)}{i+2} &= 6 \alpha_n(V)^2 \beta_n(V) \Gamma \notag \\
\tau_n(V) a_3(V) \sum_{i=1}^4 \frac{a_i(V)}{i+3} &= 4 \alpha_n(V) \beta_n(V)^2 \Gamma \notag \\
\tau_n(V) a_4(V) \sum_{i=1}^4 \frac{a_i(V)}{i+4} &= \beta_n(V)^3 \Gamma \notag
\end{align}
where $\Gamma = \frac{\beta_n(V)}{(\alpha_n(V)+\beta_n(V))^4 - \alpha_n(V)^4}$.
In practice, the system of nonlinear equations defining $a_i(V)$ must be solved numerically.  We used the \texttt{Minimize} command in Maple (Waterloo Maple Inc., Version 13) to generate a data table of values of $a_i(V)$ that solved these equations for voltage values ranging from $-20$ to 120~mV in increments of 0.01~mV.
The procedure for constructing the non-Markovian Gaussian process for the \Nap conductance is similar but slightly more complicated because there are seven time scales in the autocorrelation function.  We omit these details here and direct the interested reader to computer code available at ModelDB \cite{Goldwyn2010modeldb}.

To numerically integrate this quasistationary model, we used a forward Euler method to update the value of $V$ and $n$ in each time step, where the stochastic integral for $\eta_t$ is integrated as follows:
\begin{enumerate}
\item Compute the voltage dependent terms $\tau_n$, $\sigma_n$, $\alpha_n$, $\beta_n$, and $a_i$ using the value of $V$ from the previous time step.
\item Update the terms associated with each time scale: $A_i(t+\Delta t) = A_i(t) e^{-\frac{i \Delta t}{\tau_n}} + a_i \sigma_n r \sqrt{\Delta t} $ where $i=1,2,3, 4$, $\Delta t$ is the time step, and $r$ is a mean zero, unit variance Gaussian random generated on each time step.  
\item Update the stochastic process: $\eta_{t + \Delta t} = \sum_{i=1} ^4 A_i(t+ \Delta t)$.
\end{enumerate}



%


\end{document}